%% file: sample-authordraft.tex
\documentclass[sigconf]{acmart}

\makeatletter

\newcommand{\minor}[1]{{#1}}
\newcommand{\rr}[1]{{#1}}

\makeatother

\usepackage[utf8]{inputenc}
\usepackage{booktabs}
\usepackage{multirow}
\AtBeginDocument{%
  \providecommand\BibTeX{{%
    \normalfont B\kern-0.5em{\scshape i\kern-0.25em b}\kern-0.8em\TeX}}}

\usepackage{caption, subcaption}
\usepackage{graphicx,calc}
\newlength\myheight
\newlength\mydepth
\settototalheight\myheight{Xygp}
\setcopyright{none}
\renewcommand\footnotetextcopyrightpermission[1]{} 
\begin{document}

\title[AI-Assisted Presentation Slides Creation for Presenting Data Science Work]{Telling Stories from Computational Notebooks: AI-Assisted Presentation Slides Creation for Presenting Data Science Work}

\author{Chengbo Zheng}
\authornote{Both authors contributed equally to this research.}
\email{cb.zheng@connect.ust.hk}
\orcid{0000-0003-0226-9399}
\affiliation{%
  \institution{Hong Kong University of Science and Technology}
  \city{Hong Kong}
  \country{China}
}
\author{Dakuo Wang}
\authornotemark[1]
\email{dakuo.wang@ibm.com}
\affiliation{%
  \institution{IBM Research}
  \country{USA}  
}

\author{April Yi Wang}
\email{aprilww@umich.edu}
\affiliation{%
  \institution{University of Michigan}
  \country{USA}  
}

\author{Xiaojuan Ma}
\email{mxj@cse.ust.hk}
\affiliation{%
  \institution{Hong Kong University of Science and Technology}
  \city{Hong Kong}
  \country{China}
 }

\renewcommand{\shortauthors}{Zheng et al.}


\begin{abstract}
  Creating presentation slides is a critical but time-consuming task for data scientists. While researchers have proposed many AI techniques to lift data scientists' burden on data preparation and model selection, few have targeted the presentation creation task. Based on the needs identified from a formative study, this paper presents NB2Slides, an AI system that facilitates users to compose presentations of their data science work. NB2Slides uses deep learning methods as well as example-based prompts to generate slides from computational notebooks, and take users' input (e.g., audience background) to structure the slides. 
  NB2Slides also provides an interactive visualization that links the slides with the notebook to help users further edit the slides. 
  A follow-up user evaluation with 12 data scientists shows that participants believed NB2Slides can improve efficiency and reduces the complexity of creating slides. Yet, participants questioned the future of full automation and suggested a human-AI collaboration paradigm.
\end{abstract}



\begin{teaserfigure}
  \includegraphics[width=\textwidth]{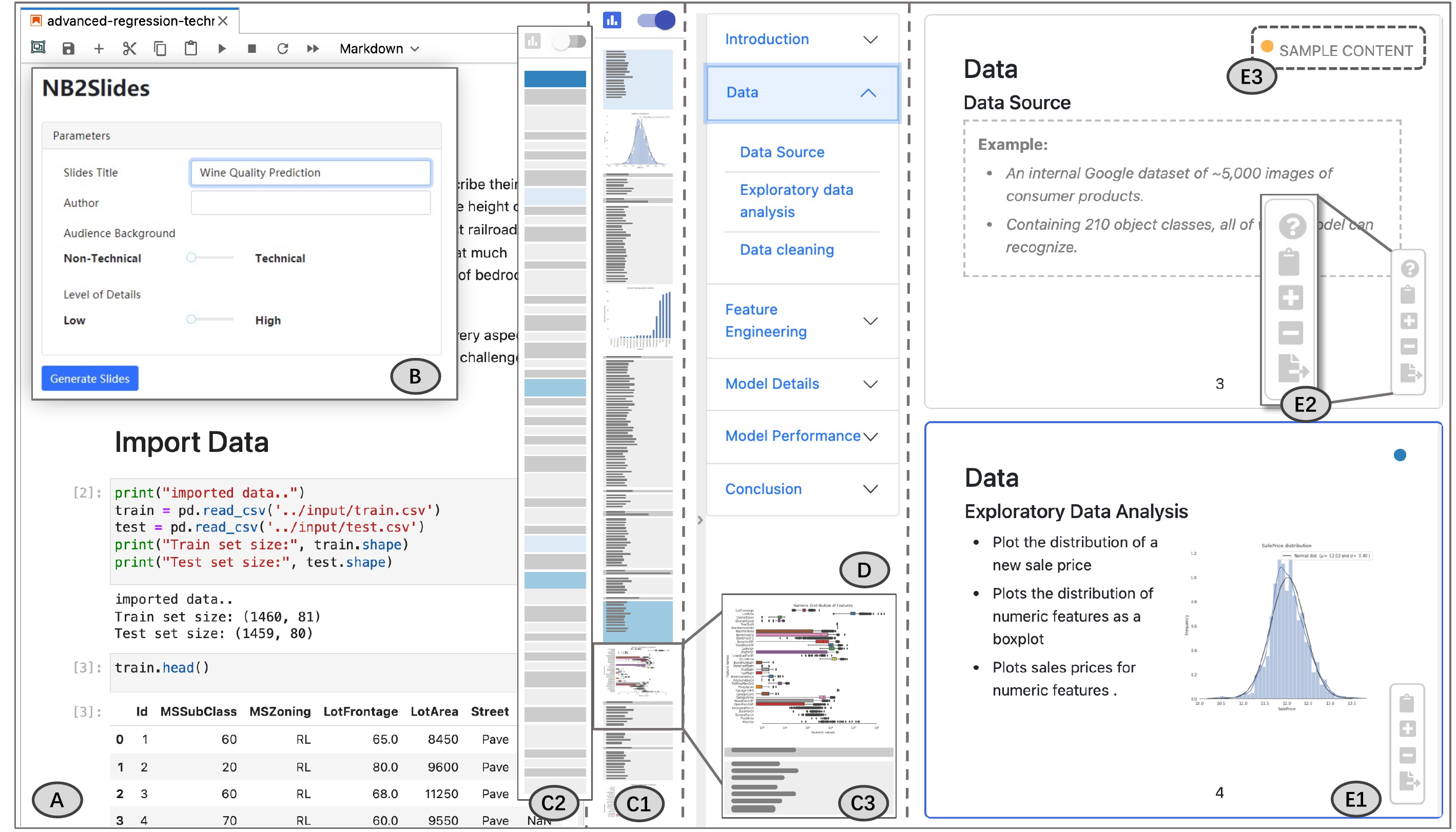}
  \caption{User Interface of NB2Slides, a human-centered AI-assisted presentation slides creation system.
  \rr{
    Users can open their jupyter notebook (A) and configure the slides generation (B). Using the notebook overview (C1-3), they can navigate through the notebook. The generated slides are presented in (E1-3) with a slides outline overview (D). The slides are either generated by deep-learning algorithms or filled with example-based prompts, as indicated by (E3). Users can modify the slides with (E2).
  }
  }
  \Description{}
  \label{interface}
\end{teaserfigure}


\maketitle

\section{Introduction}
\input{sections/01_Introduction}

\section{Related Works}
\input{sections/02_RelatedWork}

\section{Formative Study}
\input{sections/03_FormativeStudy}

\section{NB2Slides: Human-centered AI System for Presentation Slides Creation}
\input{sections/04_System}

\section{User Evaluation of NB2Slides}
\input{sections/05_UserStudy}

\section{User Evaluation Results}
\input{sections/06_Results}

\section{Discussion}
\input{sections/07_Discussion}

\section{Conclusion}
\input{sections/08_Conclusion}



\begin{acks}
We thank the feedback and comments from the anonymous reviewers that helped improve this paper a lot.
We also thank Boran Xu for contributing to the codes.
This work is supported by the Hong Kong General Research Fund (GRF) with grant No. 16203421.
\end{acks}

\bibliographystyle{ACM-Reference-Format}
\bibliography{sample-base}

\appendix
\input{sections/09_Appendix}
\end{document}

%% file: sections/01_Introduction.tex
Data science (DS) refers to the practice of using machine learning modeling (ML) techniques to generate domain-specific insights from data, then making better decisions with these data-driven insights~\cite{wang2021autods}.
Thus, DS projects are by nature interdisciplinary collaborative teamwork~\cite{Muller:2019:DSW:3290605.3300356}.
Prior works have proposed theoretical frameworks to describe a DS project's lifecycle~\cite{wang2021much,kross2021orienting}, and they found that the more technical team members (e.g., model builders) need to constantly communicate with the less technical team members (e.g., business decision makers) before, during, and after the modeling works~\cite{kross2021orienting, wang2021much, donoho201750}.
For example, after data scientists build a model, they need to present the model to the stakeholders to collect their feedback or buy-in (i.e., Model Presentation \& Stakeholder Verification in Fig.~\ref{fig:AILifecycle}), and such presentation need to be structured in less technical storytelling manner~\cite{piorkowski2021ai,mao2019data,kross2021orienting}.

However, as we may expect, the creation of such presentation is a non-trivial task. Some prior works reported that DS workers may spend hours or even days to prepare the presentation slide after they finish their core technical modeling works ~\cite{piorkowski2021ai}. 
The work is tedious and time-consuming due to multiple reasons: a) data scientists have to meticulously \textbf{locate} and distill essential information from the complex, messy, and sometimes fragmented experimental codes\cite{rule2018exploration, kross2021orienting}; b) they need to \textbf{organize} these information to construct a story narrative; and, c) they also need to consider the specific domain context and audience backgrounds, and often add additional information (e.g., visualizations, explanations, and examples) to \textbf{customize} their presentation so that it can better engage the target audience and gain their trust~\cite{kross2021orienting, brehmer2021jam, suresh2021beyond}. 


In this project, we aim to support DS workers' presentation creation task while leveraging the recent advance of AI techniques.
To motivate the problem, we conducted a formative study with seven DS workers to understand how they perform this task, and what challenges they face in today's manual process. 
Based on the study findings as well as literature review, we are convinced that DS workers desire automation solutions to help with their presentation creation work for all the three sub-tasks (i.e., locate information, organize a story, and customize it).
\rr{
But a fully-automated system is not desired --- they do not expect an AI solution can understand the value of the DS work and provide insights as they can do.
Thus, we believe a human-AI collaboration paradigm fits such tasks, which is also postulated by a previous study~\cite{wang2021much}.
AI can take over the low-level but indispensable reasoning tasks such as locating information in source files.
The human can focus on high-level reasoning activities such as digging and presenting the business value of a DS project.
}

Very recently, there are a few researchers have started the exploration of using neural-network based deep learning techniques to support various DS tasks~\cite{xin2021whither,kang2021toonnote,liu2021haconvgnn,wang2021themisto,DataRobot}.
We join their effort by designing and implementing an human-centered AI system --- NB2Slides --- to support DS workers creation of presentation slides.
NB2Slides takes a Jupyter notebook as the input, and generates draft presentation slides as the output, as notebooks and presentation slides are the most commonly used formats in DS presentations~\cite{zhang2020data,knuth1984literate, kluyver2016jupyter, Kery:2018:SNE:3173574.3173748, rule2018exploration}.
Internally, it uses a deep-learning-based approach to locate information and generate content; it also uses a template-based approach to cover the scenarios where today's AI techniques are insufficient.
The main advantage of our system is that we follow the human-centered AI design guidelines from existing literature and from our own formative study, so that our system try its best to provide a trusted, explainable, and controllable AI experience to the users, while ensuring the effectiveness of its automation solution. 
For example, a user can specify the background of the target audience, and this user customization can alternate the deep-learning language model structure thus to generate different versions of the outputs.

As a follow-up user evaluation, we designed an experiment to have 12 DS workers use NB2Slides to create presentation decks for a sample notebook. Based on quantitative and qualitative data, NB2Slides can improve participants' efficiency in creating presentation slides.
\rr{
The participants collaborated with the AI component in NB2Slides by actively refining the AI-generated draft.
Yet, participants questioned the future of full automation.
They believed automation could draft slides like what NB2Slides served. Still, human intervention is essential to uncover the business value of the models, reveal the ``secrets'' of the DS work and bring the stories out of the notebook.
Thus, they preferred our human-AI collaboration paradigm for creating slides to present DS work in the future.
}

In summary, our contributions of this paper are three-fold:
\begin{itemize}\vspace*{-6pt}
    \item We conducted a formative study to provide empirical understanding of how data scientists create presentation slides and what challenges they face;
    \item We implemented an AI-assisted slide creation system called NB2Slides to support data scientists to create presentation slides.
    \item We gathered user feedback and proposed new design suggestions and future research directions via a follow-up user evaluation study of the NB2Slides system.
\end{itemize}

%% file: sections/02_RelatedWork.tex
Our project aims to explore how DS workers create their presentation slides, and then build human-centered AI systems to support such task. Thus, in this section, we organize the literature review into three subsections: Communication in Data Science Teams, Data Science Work in Computational Notebooks, and Human-centered AI for Data Science.

\subsection{Communication in Data Science Teams}
\begin{figure}[h]
  \centering
  \includegraphics[width=0.8\linewidth]{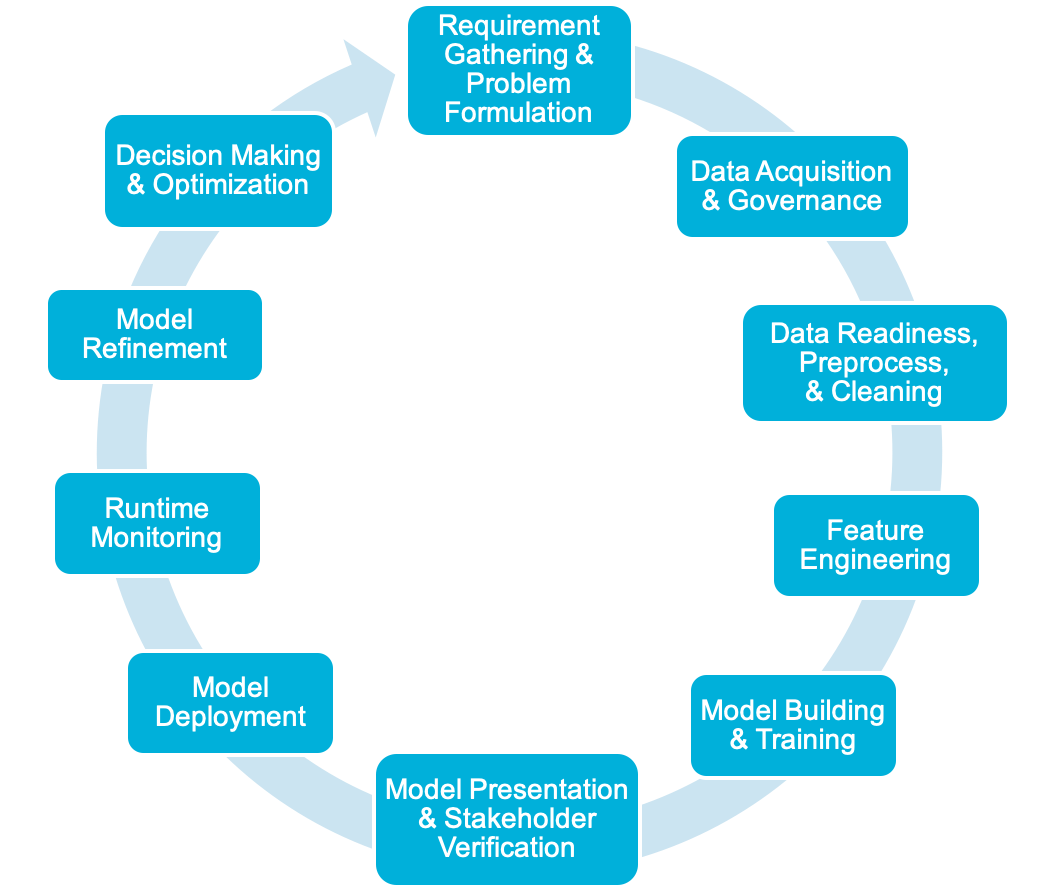}
  \caption{A 10 Stages DS/ML Lifecycle, starting at the top Requirement Gathering~\cite{wang2021autods}}
  \Description{automation model}
  \label{fig:AILifecycle}
\end{figure}

With the prominence of DS in various domains and its growing complexity, DS workers often need to collaborate with other DS workers, domain experts, and other roles~\cite{Muller:2019:DSW:3290605.3300356}.
Interpersonal communication can take place at the various stage of a DS project lifecycle in Fig.~\ref{fig:AILifecycle} ~\cite{wang2021autods}. 
For example, at the beginning of a project, an interview study with professional data scientists discovered that DS workers worked closely with clients to gain domain knowledge during problem formulation and feature engineering stages \cite{kross2021orienting}.
At the end of a project, after DS workers have done all the technical work, they need to communicate with stakeholders so that these stakeholders can make decisions based on their reports \cite{wang2021much, xin2021whither,piorkowski2021ai}.
Even during the process, sometimes DS workers need to constantly and iteratively communicate with the problem owners and domain experts to present progress, build common ground, and calibrate the next steps~\cite{mao2019data}.
Zhang et al.\cite{zhang2020data} confirmed this argument with an online survey that DS teams are characterized as extreme collaborative with high demands to communicate with a variety of stakeholders.
We begin our project with focus on the model presentation stage, as it is when primarily DS workers present their accomplished DS work to stakeholders.
In the future, we hope to generalize our methods to cover other occasions where communication happens. 

Team communication is critical to the success of a team work, but also challenging, especially for an interdisciplinary team in the DS context. 
Many existing works have investigated the team communication strategies and the communication challenges among DS teams.
For example, Mao et al. \cite{mao2019data} reported that when DS workers working with bio-medical domain experts, the differences in these two groups' motivations lead to the challenge in building ``content common ground'' and ``process common ground'', which impact the project's final outcomes.
Piorkowski et al. \cite{piorkowski2021ai} specifically focused on the inter-role communication gap with a couple of multidisciplinary DS teams as a case study. 
They identified three types of communication gaps that are common: knowledge gaps, trust building, and matching expectation.
In the public domain, Hou et al. \cite{hou2017hacking} reported an ethnography study of DS technical volunteers and subject matter experts in civic data hackathons sometimes require a third party to help ``translate'' DS workers' technical language and subject matter experts' business or domain language.
In our work, we aims to support this communication process by helping DS workers to draft their DS works in business or domain language.

In the business world, most of these interpersonal communications in a DS team are verbal presentations with the help of a well-prepared slide deck~\cite{piorkowski2021ai,xin2021whither}. 
We acknowledge that there are some other media that people use to organize their DS work (e.g., a detailed documentation~\cite{H2O,DataRobot}, a model card~\cite{ mitchell2019model}, or a model factsheet~\cite{arnold2019factsheets}), but these types of documents are more for the information archival purpose and for the government regulators to examine the DS works. 
Inside a DS team, stakeholders and DS workers still prefer the presentation decks that can easily and visually convey the high-level findings and insights~\cite{kross2021orienting}.
Therefore, we decide to focus on presentation slides deck as the target artifacts of the presentation deliverables.



\subsection{Computational Notebook as an Artifact of Data Science Technical Works}
There are a diverse range of artifacts that can represent the technical works in a DS project~\cite{wang2021much}. 
DS workers may refer to the code files, the processed data, a corresponding design requirement file, or they may sketch a ML model architecture diagram. 
In this paper, we focus on Jupyter Notebook, one implementation of the literate programming platform that millions of DS workers use for their daily work~\cite{kluyver2016jupyter,rule2018exploration, Kery:2018:SNE:3173574.3173748}. 
A typical notebook contains multiple chunks of cells mixed together, with python or R code in ``code cells'', and with rich text, plots and output results in ``markdown cells''.
DS workers love notebooks because it provides an easy to use graphical-user-interface, and it shows results simultaneously and next to the codes, thus it supports the DS work's exploration and experimentation nature~\cite{rule2018exploration}.

While DS workers find computational notebooks useful and popular, notebooks also have their drawbacks.
Rule et al.~\cite{rule2018exploration} raised the problem of the tension between exploration and explanation in computational notebooks.
Further, Chattopadhyay et al.~\cite{chattopadhyay2020s} summarized nine pain points for computational notebooks, including exploring and analyzing, managing code, sharing and collaborating, security and so on. 

\rr{
Very recently, some intelligent systems have been proposed to solve these problems.
As an example, WREX, proposed by Drosos et al.~\cite{drosos2020wrex}, can help to automatically generate readable codes in notebooks on the data wrangling.
Head et al.~\cite{head2019managing} introduce code gathering methods for notebooks to manage messy codes. 
Wang et al.~\cite{wang2020callisto} present a Jupyter extension to connect discussions around notebooks and facilitate collaborations.
All these efforts jointly improve DS workers' experience working with notebooks and help them create readable notebooks for sharing and collaboration.
}

\rr{
In our work, we decide to use the computational notebook as the representative artifact of DS workers’ technical work because it is popular and affords rich content as well as raw code.
We aim to facilitate the presentation slides creation from notebooks.
There are some existing practices for DS workers to do so.
Before creating slides, DS workers often need to clean the notebooks (especially when having technical audiences)~\cite{rule2018exploration}.
Then, they would manually migrate information from notebooks to presentation tools like PowerPoint.
But this process needs to switch between tools frequently, which is tedious and prone to error~\cite{brehmer2021jam}.
DS workers often reuse their past presentation slides~\cite{piorkowski2021ai}, which still need them to manually adapt the slides’ contents to a new notebook. 
They can also apply \texttt{nbconvert}~\cite{nbconvert}, a built-in functionality in Jupyter Notebook, to convert selected notebook contents into slideshows.
However, \texttt{nbconvert} simply copies one notebook cell to one slide with exactly the same code and outputs, but it does not summarize or generate any new content for a non-technical audience to understand.
In reality, DS workers need to digest and generate the key points from a lengthy notebook and summarize and create new content using human-readable presentations~\cite{rule2018exploration}.

To support DS workers' current presentation curation practice, we decide to leverage the recent advance of deep-learning-based AI models to summarize the key information from notebooks and organize them into a format that may be suitable to present to a non-technical audience.
With an interface built inside Jupyter Lab, DS workers can collaborate with AI to refine the generated slides and prepare the presentation.
}


\subsection{Human-Centered AI for Data Science (AutoML)}

The idea of utilizing AI automation techniques to speed up and scale up various DS works in the DS lifecyle has been popular in the last few years~\cite{wang2021much,xin2021whither}. 
People refer to this group of techniques as automated machine learning (AutoML) or automated data science (AutoDS), and in this work, we use AutoML. 
AutoML researchers have invented various AI optimization techniques to automatically select a model~\cite{he2021automl}, generate new features~\cite{galhotra2019automated}, try out hyperparameter's candidate values~\cite{liu2020admm}, and some other stages of the DS lifecycle~\cite{zhang2020data}. 
In additional, tech companies (e.g., Google Cloud AutoML~\cite{googleAutoML}), and start-ups (e.g., H2O~\cite{H2O} and DataRobot~\cite{DataRobot}) have built an entire AutoML industry sector to provide services for both technical and non-technical users to perform DS tasks and get insights from their data. 

With AutoML research and products advance quickly, some Human Computer Interaction (HCI) researchers have began to explore AutoML's user experience of and its impact on DS workers.
An interview study reported that DS workers believe that automation can give them some assistant, but it can never (and should not be designed to)  replace their jobs~\cite{wang2019human}. 
A recent experiment~\cite{wang2021much} further reveals the benefits of AutoML that collaborating with an AutoML system, DS workers can produce models with higher quality and with less human errors.

However, most of today's AutoML works have focused on the more technical DS works such as model building and feature engineering, limited attention is given to less technical but critical interpersonal communication tasks of a DS project.
A few recent information visualization research works may seem relevant for our purpose as they adopt automation techniques to support users for the data exploration tasks~\cite{wu2021multivision, wang2019datashot, shi2020calliope}.
\rr{And a more closely related work is Themisto~\cite{wang2021themisto}, where the authors utilize neural-network-based methods to generate human-readable documentation for a code snippet in a notebook so that other coders can understand the notebook better. 
But this project focuses primarily on the DS workers' technical work. Thus the target audience of the generated documentation is still the technical DS workers. 
In our project, we intend to ease the human effort of breaking the boundaries between the technical and non-technical members in a DS team by supporting the DS workers' presentation preparation effort.
Our project also differs from the NBSearch~\cite{li2021nbsearch} system, where they focused on visualization and a ``smart'' search function. The NBSearch system allows users to type a natural language query and returns the most relevant code cells. In our project, we aim to generate new contents for which non-technical DS team members may be the audience. Before the generation module, our system also needs to locate the relevant code cells for a given slide page, where the NBSearch’s semantic search algorithm inspired the design.}

We join these AutoML and HCI researchers with our ultimate goal to reveal emperical understandings and build systems to support DS teams.
Thus, we execute a three-steps research plan to 1) understand how DS workers create presentations today, 2) to implement an AI system prototype to support their work, and 3) to explore target userss feedback and design implications after they try it out.

We would like to emphasize that it is not our goal to build a ``fully automated'' system to replace human DS workers to complete the presentation creation task. We do not believe today's AI technology is capable of doing that; even if it is, we do not believe human wants that.
Instead, we would like to build an AI solution to the right level of automation, so that the outcome presentation deck of the AI system can be the starting point of human DS workers' work. Together, they can jointly and iterative improve the final outcome. 


%% file: sections/03_FormativeStudy.tex
To understand how DS workers create presentation slides today, we conducted a formative study with seven participants who have had experience in presenting their DS works. 

The formative study has two parts: we first thoroughly surveyed relevant literature~\cite{liao2020questioning, suresh2021beyond, anik2021data,kross2021orienting, mao2019data, zhang2020data,piorkowski2021ai,Muller:2019:DSW:3290605.3300356} and industrial standard documentations~\cite{mitchell2019model, gebru2018datasheets,H2O,DataRobot} to understand how a presentation deck looks like, and what contents are usually included. 
Based on these insights, we drafted an \textbf{presentation outline} to represent a common presentation structure. 
Then, we conducted a think-aloud participatory design session with seven participants. During the sessions, participants were asked to create slides decks for a given notebook
; we then showed them a pre-constructed slides guided by our outline, and asked them to reflect and co-design with us.
In the following subsections, we detail the process and key findings.

\input{tables/slide-template}

\subsection{Literature Survey and Presentation Outline}
\label{deckoutline}
\rr{
By searching with terms such as ``data science'', ``presentation'' and ``collaboration'' within human-centered DS papers published in CHI, CSCW and FAT, we surveyed prior research related to explainable AI (e.g., ~\cite{liao2020questioning, suresh2021beyond, anik2021data}), documentation for machine learning (e.g., ~\cite{mitchell2019model, gebru2018datasheets}), and data science team collaboration (e.g., ~\cite{kross2021orienting, mao2019data, zhang2020data}).
}
In addition, we investigated the outcomes of the automatic report generation feature provided by some AutoML products (e.g., H2O~\cite{H2O}, DataRobot~\cite{DataRobot}).
We found that many of these works focus on the documentation and archival of DS works, instead of communicating or presenting a DS project, thus a document is easily more than 100 pages long with too much details~\cite{DataRobot}.
We discarded those too detailed information. 
As an example, Model Card~\cite{mitchell2019model} includes information about the version of the programming framework, but such information is minimal value for a presentation~\cite{kross2021orienting}.
The initial presentation outline has five sections (\textit{Introduction, Data, Model, Model Performance} and \textit{Conclusion}) and 17 sub-sections. as shown in Table \ref{tab:slide-template} (a).

To prepare for our participatory design sessions, we used the this initial presentation outline to manually craft an example slides deck for a winning notebook solution retrieved from the HousePricePrediction challenge on Kaggle\footnote{\url{https://www.kaggle.com/lavanyashukla01/how-i-made-top-0-3-on-a-kaggle-competition}}~\cite{kaggle}. 
Kaggle is a platform for organizations to publish data challenges and many data scientists submit corresponding solutions on it. 
This notebook has three main sections: (1) Exploratory data analysis(EDA), which corresponds to the \textit{Data} section in our outline; (2) Feature Engineering, corresponding to the section under the same name; (3) Model Training, corresponding to the \textit{model details} and the \textit{model performance} sections. 
In addition, in order to create \textit{introduction} and the \textit{conclusion} sections according to our outline, we had to look for additional information (i.e. wikipedia) beyond the notbeook. 
We attach this example presentation slides in the supplementary material.

\subsection{Participants in Participatory Design Sessions}
We recruited seven participants who have DS background by posting our recruitment message on online social media and our professional networks. In the end, eight participants (three women and five men) joined, but one participant failed to create her own slides due to technical difficulties, thus we exclude her from our analysis (Table~\ref{tab:formative}). 

\subsection{Procedure of Participatory Design Sessions}
The participatory design sessions were conducted via online Zoom meetings, which lasted between 40 minutes to 1 hour. 
All sessions were recorded for subsequent analysis with the participants' consent.

\pagebreak

In each session, participants have two tasks: first, we would like them to follow their existing practice to create a slides deck for a given notebook, from which we observe how they do it, and what final slides look like; second, we share the pre-crafted slides mentioned in Sec.~\ref{deckoutline}, and invite them to reflect on their own version of the slides with ours.
 
Given the notebook is fairly simple, and based on our pilot study, we gave participants 15 minutes to read the notebook and create a slides deck. 
The target communication scenario was to present this DS project to a real estate client team (with both business and technical members). 
We suggested that they focus on the content rather than styling and formatting.
During the session, we asked them to think-aloud.

We then prompted them with our prepared example slides deck, and asked them to co-design this deck together with us. They may also refer back to their own deck in this practice. 
During this process, we also asked semi-structured interview questions specifically related to the challenges that they face, and the potential automated solutions for them.

\subsection{Results of Participatory Design Sessions}
\rr{
We transcribed the recordings of all sessions, coded the transcripts following the thematic analysis ~\cite{braun2019reflecting}, and analyzed the artifacts participants created by comparing them to our example slide deck.
}
\rr{
All participants agreed that the given notebook is similar to their working ones and they created the decks following their current practice.
They also praised our design of the participatory design session as they felt it quite realistic to their day-to-day work.
}
Among seven participant-created presentation decks, six have a similar structure to our example slides, which suggested the validity of our presentation outline.
One participant (P7) skipped the \textit{data} and \textit{feature engineering} sections and directly go to the \textit{model} section after stating the problem and goal in the slides.

\subsubsection{Challenges and Opportunities for Presentation Slides Creation}
\label{sec:techniques}
Almost all participants mentioned that one particular challenge of creating presentation slides is to customize for different target audiences.
Participants need to understand their audience: What they can understand, what they care about, and what they are attracted to.

\begin{quote}
    ``Some stuff may be understood by me but not by the audience''-P03 
\end{quote} 

\rr{
These DS workers need to frequently switch between their technical materials (e.g., notebook) and the draft slides to select the information to present according to the intended audience.
This process is often found challenging.
}
\begin{quote}
``The steps are relatively long and it is hard to find the key points'' -P01

\rr{
``It’s hard to find the [output] image that should be used''-P07}
\end{quote}

\input{tables/formative}

They would definitely hide the technical details for non-technical audience, or leave those details into an appendix section in case someone may ask. P06 further elaborated:
\begin{quote}
``It is hard to clarify these [technical details] clearly. The audience also doesn't want to know about these. They prefer to know the final results.''-P06
\end{quote}
Besides, ~\cite{kross2021orienting} reported that there is a knowledge transfer from data scientists to clients in which DS terminologies are framed into business terms. 
Our findings echo this insight.
P6 said instead of showing the feature list, ``\textit{you must tell them what these features are depicting}''.
Participants also noted the importance of showing the business value of the model.
P07 reported that they really need to show ``\textit{how much benefit the model may bring to the company}''. 

\rr{
The participants also demonstrated the techniques they often used in presenting DS works, including comparing with traditional strategies (P02, P06, P07), selecting data points for case analysis (P03, P06), and transforming metrics from accuracy to human labors or budgets (P07).
}

\subsubsection{Co-Design of the Presentation Outline}
\label{sec:proscons}
Participants in general appreciate our design of the presentation outline. 
In particular, most of them (6 out of 7) liked the \textit{introduction} and \textit{conclusion} sections. 

Participants believe even if they did not follow the exact outline to create a presentation deck, some parts of the outline serve as a reminder to them.
For example, although none of participants discuss \textit{model interpretability} in their own slides, but when seeing that section in our outline, they all expressed the importance of such information when communicating with audience.


In addition, some participants expressed their concerns regarding a few aspects of our outline, mostly still around the audience background.
Some believe it still has too much technical details which will confuse the non-technical audience, although these details are recognized to be important and needed by other participants. 
As P03 said, ``\textit{I also thought about explaining the model, but I don’t think they can understand what I said}''. 


\subsubsection{``Some automation could be helpful.''}
\label{sec:needsautomation}
When being asked whether they see opportunities for automated solutions to help them with this presentation slides creation process, all participants agreed that they want some automation. 
However, their desires of automation differ across the different sections of the outline.
Participants certainly hope the technical details in a notebook can be automatically extracted and summarized into the slides. 
\begin{quote}
``The EDA part can be put into the slides by the system automatically. If it is too much, I can just delete redundant points. Feature engineering and the model details can be done similarly.''-P03.
\end{quote}

P06 added that AI-generated contents needs to have explanations: ``\textit{I want to know how your information came from}''.

Not surprisingly, participants believed that there are certain types of contents can never be automated by AI (e.g., the introduction and the conclusion), as they require external and domain information beyond the notebook.
For these sections, participants prefer to do by themselves. 
The reasons may be they lack confidence in today's ML algorithms, as P02 said that ``\textit{AI technologies, such as generating resumes, are not 100\% satisfactory}''.
As an alternative, participants hope that we can give some ``good examples'' or ``best practices'' as part of the outline so that they can reference on what they should add into these sections: 
``\textit{I would rather be able to fill in the blanks}''(P02).

\subsubsection{Summary of Design Requirements}
In summary, our participants praised the effort of creating a presentation outline, they helped us iteratively refined the outline, and expressed their interests of having automation to help with some parts of the presentation but not all. 
\rr{
They suggested that good presentation slides for a DS project should be:
\begin{itemize}
    \item Effectively summarizing the key information about the data and the models used.
    \item Conveying the insights in a language that fits the audience's background (e.g., technical or non-technical).
\end{itemize}
}

Based on these findings, we summarize the following design requirements for a human-centered AI system to support presentation slides creation for DS workers:

\begin{itemize}
    \item \textbf{DR1: The AI system should customize output presentation slides for the different backgrounds (e.g., technical and non-technical) of the target audience.} 

    \item \textbf{DR2: For those information exist in a DS project (e.g., ML codes in a notebook file), the AI system should accurately and automatically extract and summarize them into presentation slides.}
    
    \item \textbf{DR3: For those AI-generated slides, the AI system should provide explanations on how these slides are generated.} 
        
    \item \textbf{DR4: For those sections that the AI system can not automatically generate, it should also provide some good examples or best practices to help users to write them on their own.}

    \item \textbf{DR5: The AI system should allow users to refine those AI-generated slides}
\end{itemize}

%% file: tables/slide-template.tex
\begin{table*}
    \renewcommand{\arraystretch}{1.02}
        \centering
        \begin{tabular}{ll} 
        \begin{tabular}{ll} 
         \toprule
         Section                      & Subsection               \\ \midrule
         \multirow{2}{*}{Introduction}      & Purpose and Intended use          \\
                                            & Workflow                          \\ \midrule
         \multirow{4}{*}{Data}              & Data Source                       \\
                                            & Exploratory Data Analysis         \\
                                            & Data Cleaning                     \\
                                            & Feature Engineering               \\ \midrule
         \multirow{5}{*}{Model}             & Model Input                       \\ 
                                            & Model Output                      \\ 
                                            & Optimization Goal                 \\
                                            & Model Alternatives                \\
                                            & Model Details                     \\ \midrule
         \multirow{3}{*}{Model Performance} & Metrics                           \\ 
                                            & Performance                       \\ 
                                            & Model Interpretation              \\ \midrule
         \multirow{3}{*}{Conclusion}        & Suggestions                       \\
                                            & Ethical \& Legal considerations   \\
                                            & Limitation \& Risks               \\
        \bottomrule
        \multicolumn{2}{c}{\textbf{(a)\ \ The slides outline for ``\textit{technical}'' audience}}\\
\label{tab:dimFFT}
        \end{tabular}
        &\begin{tabular}{ll} \toprule
         Section                            & Subsection                        \\ \midrule
         \multirow{3}{*}{Introduction}      & Purpose and Intended use          \\
                                            & Workflow                          \\ 
                                            & Data Source                       \\ \midrule
         \multirow{3}{*}{Model}             & Model Input                       \\ 
                                            & Model Output                      \\ 
                                            & Optimization Goal                 \\ \midrule
         \multirow{3}{*}{Model Performance} & Metrics                           \\ 
                                            & Performance                       \\ 
                                            & Model Interpretation              \\ \midrule
         \multirow{3}{*}{Conclusion}        & Suggestions                       \\
                                            & Ethical \& Legal considerations   \\
                                            & Limitation \& Risks               \\ \midrule
         \multirow{3}{*}{Appendix: Data}    & Exploratory Data Analysis         \\
                                            & Data Cleaning                     \\
                                            & Feature Engineering               \\ \midrule
         \multirow{2}{*}{Appendix: Model}   & Model Alternatives                \\
                                            & Model Details                     \\
        \bottomrule
        \multicolumn{2}{c}{\textbf{(b) \ \ The slides outline for ``\textit{non-technical}'' audience}}\\
        \label{tab:dimGMM}
        \end{tabular}
    \end{tabular}
    \vspace*{0.5em}
    \caption{The presentation template outline we used in NB2Slides and the formative study. (a) is also used in the formative study.}
\end{table*}
\label{tab:slide-template}

%% file: tables/formative.tex
\renewcommand{\arraystretch}{1.2}
\begin{table}[H]
    \centering
    \begin{tabular}{@{}lll@{}} \toprule
     PID & Gender & Job Roles in Data Science \\ \midrule
     P01 & Male     & Citizen Data Scientist \\ 
     P02 & Male     & Citizen Data Scientist \\
     P03 & Male     & AI-Ops/ML-Ops \\
     P04 & Male     & AI-Ops/ML-Ops \\
     P05 & Female   & Expert Data Scientist \\
     P06 & Male     & Expert Data Scientist \\
     P07 & Female   & Expert Data Scientist \\
    \bottomrule
    \end{tabular}
    \caption{Demographics of the participants in the formative study. }
    \label{tab:formative}
\end{table}

%% file: sections/04_System.tex
Based on the design insights from the formative study and from other relevant works, we implemented NB2Slides, an human-centered AI system to support DS workers to create presentations using their computational notebooks.
The NB2Slides system takes in a user's notebook and their user configurations, automatically generates slides from the notebook using either a neural-network-based ML approach or use an example-based approach, renders the generated slides side-by-side with the original notebook for users to explore, and allows users to further perform refinement.

NB2Slides consists of a server-side backend for processing notebook content and generating slides, and a client-side user interface that is implemented as a Jupyter Lab plugin.


\subsection{Slide Generation Backend}
\begin{figure*}[h]
  \centering
  \includegraphics[width=0.9\linewidth]{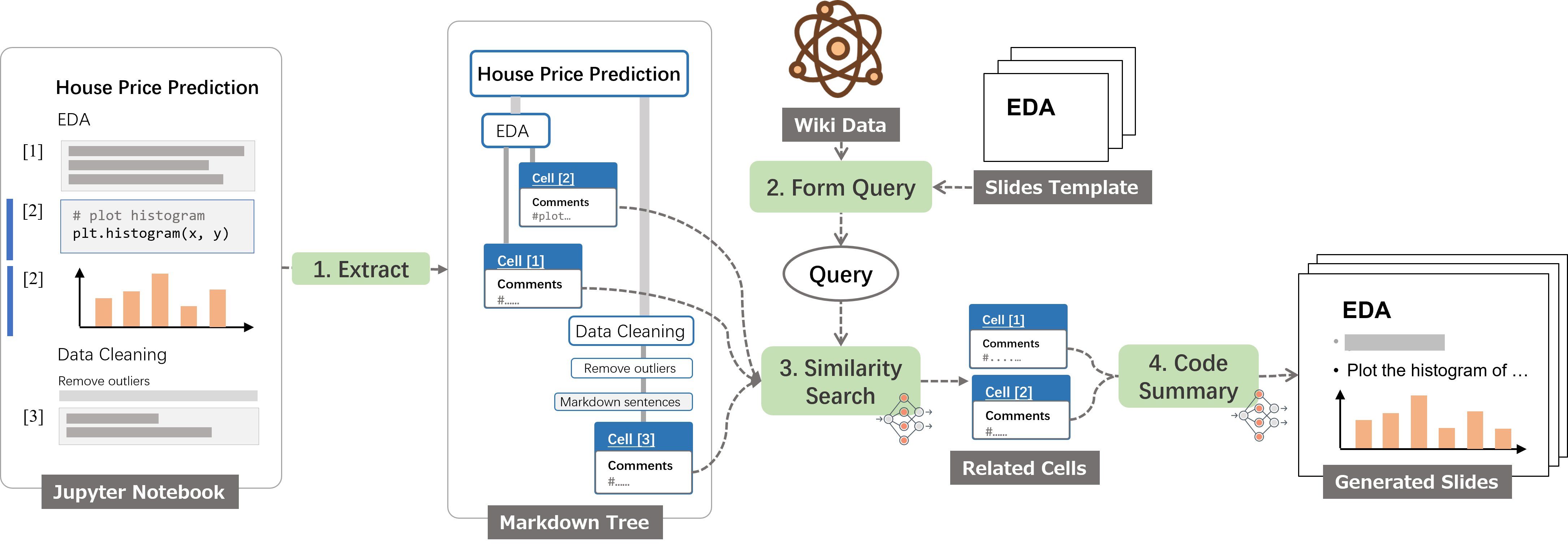}
  \caption{The NB2Slides system's slides generation backend.}
  \Description{automation model}
  \label{model}
\end{figure*}

In this section, we present how NB2Slides generate slides from a notebook using either a neural-network-based ML approach or use a heuristic example-based approach. 
\rr{
The system architecture is illustrated in Fig.~\ref{model}. From left to right, the system 1) reads in a Jupyter notebook and extract its structure; 2) it also reads in our crafted slides outline template; 3) it then queries the notebook with the template to locate related cells from the notebook for each of the outline template section; and lastly 4) it generates the slides with \textit{bullet points} and \textit{output figures} with a neural-network model.
}
 
Worth to mention, after the system reads in and parses the notebook content, depends on the user configuration (e.g., ``non-technical'' or ``technical audience''), NB2Slides can choose the two different versions of the slides outline shown in Table.~\ref{deckoutline} to satisfy user's design requirements (\textbf{DR1}).
For example, ``\textit{EDA}'', ``\textit{Data Cleaning}, ``\textit{Feature Engineering}'', and ``\textit{Model Alternatives}'' are considered to be too detailed for a non-technical audience based on the participants' feedback from the formative study, thus they will be left into appendix.

For those more technical slide sections such as ``\textit{EDA}'', ``\textit{Data Cleaning}'' and ``\textit{Feature Engineering}'', they often have corresponding code cells in the notebook.
If that is true, NB2Slides automatically composes slides for such topics by extracting relevant code cells from the notebook and then summarizing them into bullet points using natural language processing (NLP) methods. 
\rr{
Inspired by the semantic search mechanism of NBSearch\cite{li2021nbsearch}, NB2Slides first applies language models to embed queries of topics and notebook cells into vectors.
Then for each topic, it locates relevant code cells by similarity search based on distances between the topic's query vector and associated vectors of all code cells. 
For each code cell, we take not only the vector of itself but also vectors of its related markdown cells into consideration.
With this similarity-based search, NB2Slides matches the slides template with the notebook globally without asking specific language or order requirements.
After relevant code cells are located and matched with slides sections, NB2Slides applies a pre-trained deep neural network~\cite{elnaggar2021codetrans} to summarize the codes and corresponding documentation into human-readable sentences.
}
This deep-neural-network approach aims to satisfy \textbf{DR2}.

For replication purpose, here we briefly introduce how the NB2Slides automated-slides-generation pipeline is implemented with step 1 to 4 in Fig.~\ref{model}:
\begin{itemize}
    \item Step 1. NB2Slides builds a tree-structure based on the hierarchical structure of cell contents in the notebook. 
    \rr{
    A markdown cell with a header will be the child node of another markdown cell with a header at a higher level preceding it. 
    If there is only one header at the highest level, it becomes the root of the Markdown tree. Otherwise, we create a fake root as the parent of all markdown cells with the highest-level headers.}
    Markdown text that is not a header will be the child node of its immediate header. All code cells are leaf nodes. Comments in code cells are included in the corresponding leaf node.
    
    \item Step 2 \& 3. After all notebook contents are organized into the tree-structure, we then compute the cosine similarity between the tree-node's embedding, and the slides section description's embedding.
    All words are segmented into sentences using the punkt tokenizer in NLTK\footnote{https://www.nltk.org/\_modules/nltk/tokenize/punkt.html}. The sentences are then fed into the supervised SimCSE~\cite{gao2021simcse}, a state-of-the-art sentence embedding model, to compute the embedding vectors. We rank all leaf nodes using the similarity score, and pick the top $k$ cells for generating slides content (Fig.~\ref{model}(4)). The $k$ value of most queries is 3. Some other queries (e.g., data source) are given $1$. 
    \rr{If a leaf node is in the top-$k$ lists of multiple slides sections, it will be assigned to the closest section.
    Using this approach, we can also find the relevant \textit{outputs} (e.g., plots, tables) and paste it into the corresponding slide.}
    
    \item Step 4. NB2Slides uses a t5-based code summarization model in CodeTrans~\cite{elnaggar2021codetrans} pre-trained on python codes to generate the final contents.
    The minimum length of the inferred sentences is determined by the level-of-details parameter pre-set by users.
\end{itemize}

Our formative study has suggested that for certain sections of the slides, today's AI method is incapable of generation. Also, users prefer to fill in those contents themselves, such as \textit{Purpose and Intend Use}, \textit{Suggestions} and \textit{Ethical \& Legal consideration} (\textbf{DR4}).
Hence, to generate these slides deck sections, NB2Slides places the pre-crafted examples and How-Tos to prompt users to insert proper and individualized contents themselves. 
These examples and How-Tos are the result of our co-design session with participants, inspired by model cards\footnote{https://modelcards.withgoogle.com/face-detection}.

\subsection{Human-centered AI System's UI Frontend}
\label{sec:interface}

On the client-side, we design NB2Slides user interface (shown in Fig.~\ref{interface}) as a plugin for Jupyter Lab for a wide adoption. 
The system has a notebook panel (A), a configuration panel (B, and disappear after user selection), a visualization navigation column (\textit{notebook overview}) (C-1 \& C-2), an outline structure (\textit{slides outline overview}) (D), and a rendered presentation slides panel (E).

The user triggers NB2Slides by a button in the menu bar, then the system opens a configuration panel (Fig.~\ref{interface}B). 
Users can fill in meta-information about the presentation and specify the background of the intended audience as well as the expected level-of-details of the generated slides (Fig.~\ref{interface}B).
These configurations are then sent to backend and make effects on the outline selection and model generation hyperparamters, this is to satisfy \textbf{DR1.}

Then, for sections such as ``\textit{EDA}'' of which the content can be found in relevant notebook cells, the server automatically constructs bullet points in corresponding slide(s) using deep learning models (\textbf{DR2}). 
\rr{
The points are sorted by order of their corresponding cells in the notebook.
}
For remaining sections without matched cells such as \textit{Introduction}, the system instead fills in examples-based prompts, and we designed a special indicator for it (\ref{interface} E-3) (\textbf{DR4}).

\begin{figure*}[h]
  \centering
  \includegraphics[width=0.9\linewidth]{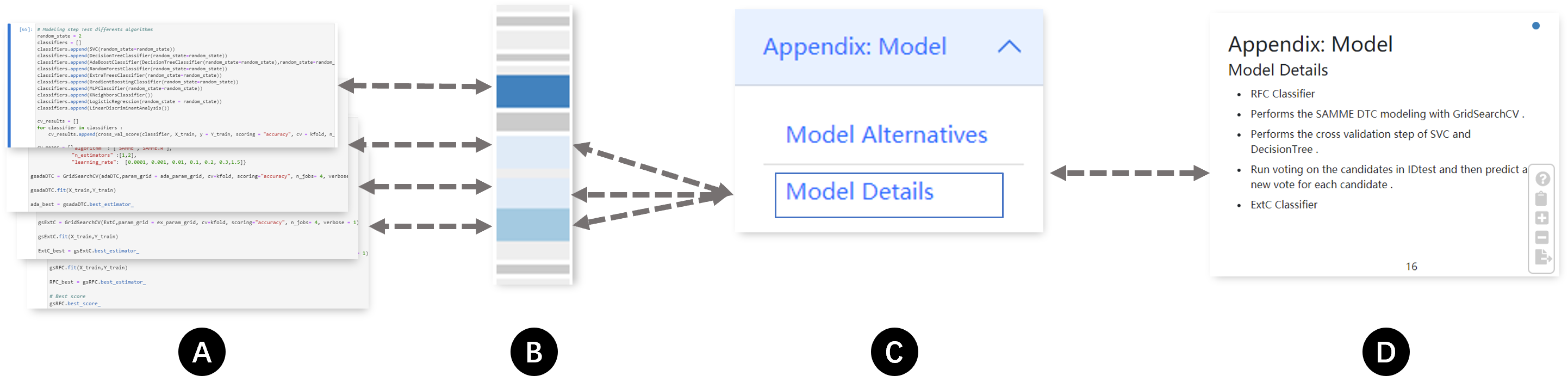}
  \caption{Illustrate how NB2Slides provide explanations by linking the generated slides. 
  }
  \Description{formative study}
  \label{linking}
\end{figure*}

For those automatically generated contents, NB2Slides supports interactive linking from a slide to associated notebook cells:
\textit{slides outline overview} (\ref{interface}D) that summarizes the slides with a table of contents; and \textit{notebook overview} (\ref{interface}C) that visualizes the notebook.
\rr{
By selecting a slide in \textit{slides outline overview} or slides panel, the relevant cells are highlighted in \textit{notebook overview} (\ref{linking}B) with the degrees of shades encode their similarity scores (computed by step 3 in \ref{model}).
\textit{Notebook overview} can dynamically link to the original notebook through clicking.
}
This process is better illustrated in \ref{linking} to support users understand the explanation of the automatic generation process (\textbf{DR3}).
\rr{
Besides, this interactive linking compensates for the possible inaccurate code summarization by allowing users to trace the source codes, which supports users to easily explore or refine the generated content (\textbf{DR5}).
}

\rr{
Users can add, delete or download slides using the tool panel (\textbf{DR5}) (Fig.~\ref{interface} E-2). 
They can also modify the text in the slides using Markdown-like grammar and copy outputs like plots from notebooks to the slides to enrich the content.
}
While some of the above features are better supported in commercial products like Microsoft Powerpoint, and users can certainly download the drafted deck and use other tools to edit, but NB2Slides facilitates users to refine the slides by connecting with the original notebook. 
More features can be added to NB2Slides to help users modify artifacts based on the notebook, for example, sync cells with slides~\cite{head2020composing}. We leave these features for future work. 


\rr{
\subsection{Input Requirements}
\label{sec:inputs}
NB2Slides can generate presentation slides for any Jupyter notebook.
But, the system can generate much better output slides, if the input notebook satisfies the following three input requirements:

\begin{itemize}
    \item[(\textbf{IR1})] Few excess code cells. 
    If the input notebook has too many excess, redundant or orphan codes,  the system may mistakenly take these codes as the sources for the generation model to generate slides.
    \item[(\textbf{IR2})] High-quality documentation. If the input notebook has poor or little documentation, the system’s code-locating module may locate the wrong cells, which hurts the accuracy of the generated slide’s content.
    \item[(\textbf{IR3})] A complete DS workflow. The system uses an outline template covering every DS lifecycle stage (Fig. \ref{fig:AILifecycle}). If the input notebook does not have codes related to ``EDA'' or ``Feature Engineering'', the generated corresponding slides will have empty content.
\end{itemize}

The three input requirements (\textbf{IR}) are nice-to-have but not mandatory requirements. We believe these requirements are not that hard to satisfy and should not impose much extra work for DS workers for the following three reasons:
First, when DS workers plan to share their notebooks and collaborate with their colleagues, they are very likely to clean those excess code cells in the notebook~\cite{rule2018exploration} thus satisfy \textbf{IR1}.
Second, despite the fact that DS workers may neglect to document their codes,  there are more and more automated tools to assist DS workers in effectively adding documentation to notebooks~\cite{wang2021themisto}, which helps to satisfy \textbf{IR2}.
Third, even if a notebook does not have certain stages of a DS lifecycle (e.g., data cleaning or feature engineering), NB2Slides is still able to generate a complete slides presentation. DS workers only need to remove the empty slides representing those missing stages.
In sum, more and more researchers are working on automated solutions to improve a notebook’s readability and documentation quality. Thus, we believe these three Input Requirements will be easier to satisfy in the future.

As a part of our evaluation of the system's robustness, we randomly sampled 21 notebooks from Kaggle and prior research papers~\cite{wang2021themisto, head2019managing}, and tested them with the system. 
NB2Slides successfully generates slides for all these notebooks.
But for the generated slides, some sections may be missing~\footnote{The reason can be either NB2Slides failed to extract the corresponding content or the original notebook does not contain these contents.} and some bullet points are positioned in the wrong section. 
\ref{tab:eval} presents statistics of the 21 generated notebooks.
}
We also present four examples of generated slides (two successful cases and two less successful cases) in the Appendix.
\input{tables/evaluation}%

%% file: tables/evaluation.tex
\begin{table}[H]
    \centering
    \begin{tabular}{@{}lcc@{}} \toprule
     Section & \# occurrence & Avg. error rate \cr \midrule
     EDA & 21     & 27.8\% \cr 
     Data Cleaning & 14     & 33.3\% \cr
     Feature Engineering & 10     & 25.0\% \cr
     Model Input & 21 & 47.6\% \cr
     Model Output & 17 & 35.3\% \cr
     Model Performance & 20     & 24.2\% \cr
     Model Details & 20   & 16.5\% \cr
    \bottomrule
    \end{tabular}
    \caption{\minor{The number of occurrence of each section in the 21 slides generated from the sampled notebooks and the percentage of the bullet points that were incorrectly extracted to each section.}}
    \label{tab:eval}
\end{table}

%% file: sections/05_UserStudy.tex
We conducted a follow-up user evaluation with 12 DS workers to ask their use NB2Slides to create a slide within an experimental setting. The study received institutional IRB approval and was conducted online through Zoom meetings. 
Through this study, we aim to answer 1) how do users use such a system in practice? and 2) what are their general attitudes towards the AI-based systems in supporting DS works.

\subsection{Participants}
We recruited 12 data scientists (6 female, 6 male) through our social network and word-of-mouth. 
To be eligible for the study, \nobreak participants need to self report prior experience in carrying out and presenting data science projects, and familiar with Python and Jupyter notebooks.
According to the background survey results(Table.~\ref{tab:participant})

\input{tables/participants}

\subsection{Experiment Procedure and Task}
We deployed our system on a cloud server, after participants joined the system and gave consent, we shared  a task sheet PDF file which contains the link to the system, the system tutorial, the tasks they need to perform, and a link to the post-study survey. 
The study procedure was designed as follow: 1) participants have 10 minutes to get familiar with NB2Slides with a sample notebook and with the guidance of the tutorial; 2) then, they have 5 minutes to get familiar with the actual experiment notebook; 3) they have 20 minutes to use the system to create a slides deck; 4) once they finish the task, we ask them to fill up the post-study survey and follow with a semi-structure interview. 
In total, one session lasts about 1 hour.
These time considerations are tested and refined using three pilot study sessions before the actual experiment starts.


The task we gave participants was straightforward: ``You need to spend the next 25 minutes to prepare a slide deck for a 10-min presentation, including Q\&A (the presentation won't happen in this study). This 10-min presentation is oriented towards a client team with both business(e.g., CEO) and technical audience (e.g., CTO). 

\rr{
The experiment notebook, consisting of 19 code cells and similar to highly-voted notebooks on Kaggle, represents a typical, complete DS workflow. 
It uses the UCI Red Wine dataset~\cite{cortez2009modeling} with a goal to build models to predict the red wine quality. 
The experiment notebook combines a few winning notebooks from the Kaggle challenge\footnote{\url{https://www.kaggle.com/vishalyo990/prediction-of-quality-of-wine}}\footnote{\url{https://www.kaggle.com/solegalli/create-new-features-with-feature-engine}}\footnote{\url{https://www.kaggle.com/niteshyadav3103/red-wine-quality-classification}}\footnote{\url{https://www.kaggle.com/vipin20/step-by-guide-to-predict-red-wine-quality-eda}}, each of which focuses on one or two different stages in a DS workflow.
\minor{
We extracted codes and documentations from these notebooks to form the final four-section notebook (“EDA”, “Data Preprocessing” “Feature Engineering”, and “Models”).
We did some minor edits to make the final notebook complete and easy for participants to read:
\begin{itemize}
    \item Some cells are omitted to avoid being verbose and repetitive.
    \item Some cells were merged or split for clearness. For example: classifiers of the final notebook were placed in one single cell while the source notebooks had each classifier in an individual cell.
    \item Some codes that are commonly used in data science, as well as Kaggle competitions were added. For example: computing F1 scores of models; plotting the cross-validation scores.
\end{itemize}
}
}
We also provided a short introduction of the background of the UCI Red Wine dataset\cite{cortez2009modeling} and the purpose of creating the prediction model at the beginning of the notebook.
\minor{
The final notebook can be found in the supplementary material.
}


\subsection{Measures}
We collected five types of data from the study: 
\begin{itemize}
    \item user behaviour data coded from video recordings with participants' consent;
    \item system server logs data, such as users selecting audience background, modifying a slide, adding a slide, deleting a slide with timestamps;
    \item the final slides deck;
    \item post-study questionnaire;
    \item post-study semi-structured interview;
\end{itemize}
The first three types of data are to answer the question 1) how do users use such NB2Slides in practice; and the last two types of data are to answer the answer 2) what are their general attitudes towards the AI-based systems in supporting DS works?



The post-study questionnaire has three parts: (1) user perceptions to the final slides in four dimensions, i.e., overall satisfactory, clearness, attractiveness, and completeness; (2) user perceptions of the following tasks in the creation process, namely, slide authoring as a whole, constructing the outline, locating the information, gathering the information, and customizing the slide deck for the audience; (3) user perceptions of NB2Slides regarding its usability, accuracy, trust, satisfaction, and adoption propensity based on ~\cite{wang2021themisto, weisz2019bigbluebot}.

In the semi-structured interviews, we asked the participants about their experience, reflections, and attitudes towards NB2SLides or similar AI-based systems. Example questions are such as:
\begin{itemize}
    \item ``What do you think of the NB2Slides system?''
    \item ``How did you take the audience's background into consideration when creating the slides?''
    \item ``What possible improvements can be made to our system?''
    \item ``How do you think about the future of having more automation help in DS works''
\end{itemize}

We used the automated transcription service provided by Zoom to convert the interview recordings to text, reviewed the results in comparison to the original audio, and corrected the errors in transcription. We followed the steps of the reflexive thematic analysis~\cite{braun2019reflecting} to code and develop themes from the transcripts. 
The identified themes include ``Customization after generating the slides'', ``Automation provides a good starting point'', ``Human ability outside automation scope'', and so on. 
We discuss our findings based on the discovered themes.
The coding book can be found in the supplementary material.

%% file: tables/participants.tex
\renewcommand{\arraystretch}{1.2}
\begin{table*}
    \centering
    \begin{tabular}{@{}lllll@{}} \toprule
     UID & Gender & Job Roles in Data Science & Frequency on DS/ML Presentation & Python Skill \\ \midrule
     U01 & Female & AI-Ops/ML-Ops & Frequently & Very Good \\ 
     U02 & Female & Expert Data Scientist & Almost Always & Excellent \\
     U03 & Male & Expert Data Scientist & Frequently & Good \\
     U04 & Female & Expert Data Scientist & Very Frequently & Fair \\
     U05 & Female & Expert Data Scientist & Frequently & Very Good\\
     U06 & Female & Citizen Data Scientist & Frequently & Fair \\
     U07 & Male & Expert Data Scientist & Almost Always & Fair \\
     U08 & Male & Expert Data Scientist & Occasionally & Very Good \\
     U09 & Male & Expert Data Scientist & Rarely & Good \\
     U10 & Male & AI-Ops/ML-Ops & Frequently & Good \\
     U11 & Female & Expert Data Scientist & Occasionally & Very Good\\
     U12 & Male & AI-Ops/ML-Ops & Frequently & Excellent\\
    \bottomrule
    \end{tabular}
    \caption{Demographics of all the participants. }
    \label{tab:participant}
\end{table*}


%% file: sections/06_Results.tex
In this section, we present the user evaluation results.
We first present how participants use and perceive the usefulness of NB2Slides. 
Then we report the human-AI collaborative slide creation paradigm enabled by the NB2Slides system, and how participants interpret the potential future of DS works.

\begin{figure*}
  \centering
  \includegraphics[width=0.9\linewidth]{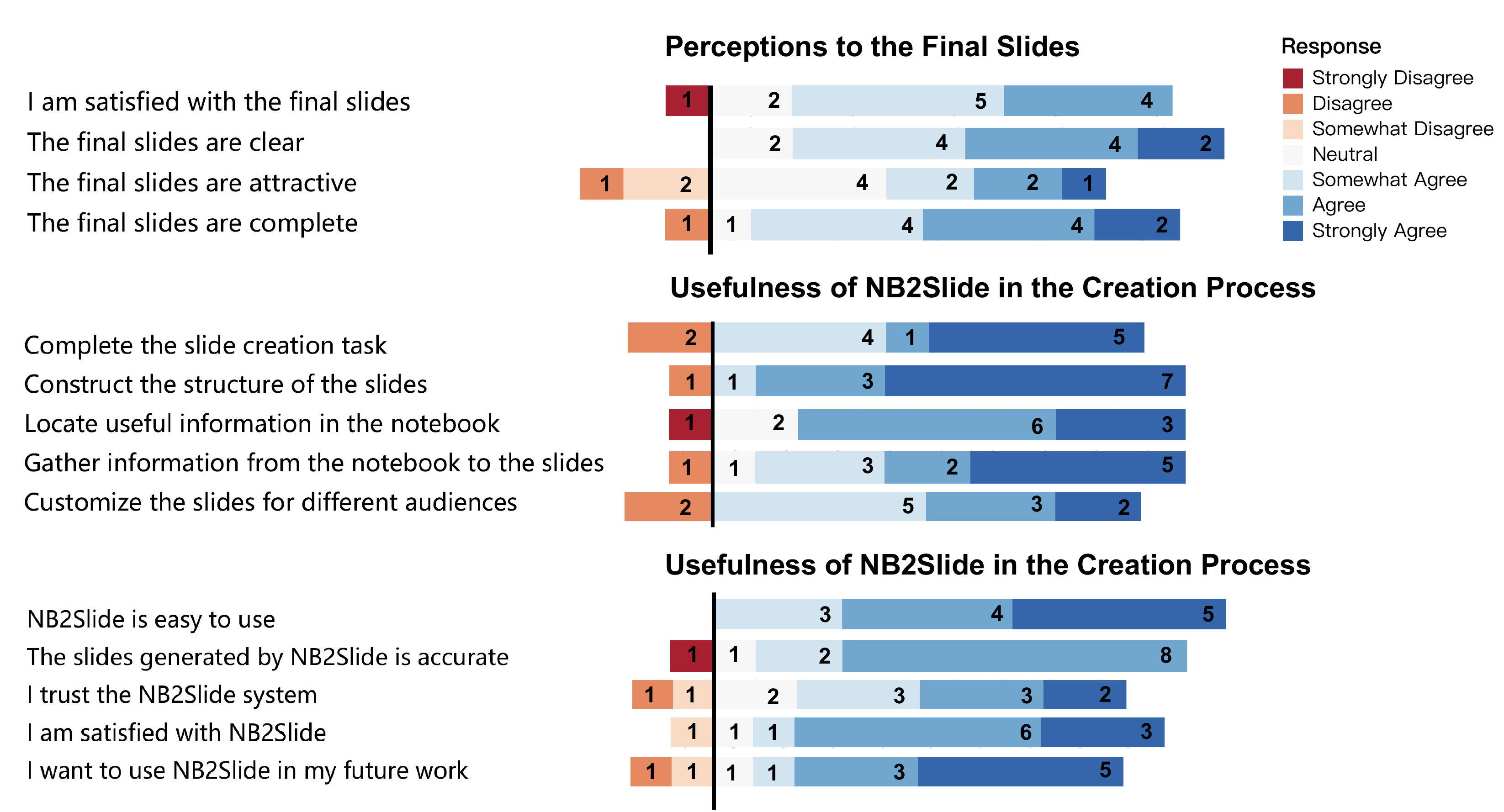}
  \caption{\label{fig:questionnaire} Results of the post-study questionnaire
  }
\end{figure*}

\subsection{How Did the Participants Leverage the AI Assistance of NB2Slides?}
\label{sec:usability}
All participants successfully completed the slide creation within 20 minutes. Overall, the post-study questionnaire reveals that the participants is leaning towards positive about the quality of the final slides, the creation process, and the NB2Slides system(Fig.~\ref{fig:questionnaire}). 

\subsubsection{Automation provides a good starting point}
\label{sec:result1}
After giving parameters like audience background, NB2Slides can generate structured slides with (1) automated generated contents mixed with text and figures, and (2) example-based prompts.
Participants liked the generated slides mainly for the following reasons.

\rr{

First, participants appreciated how the generated slides are organized. 
Most participants agreed that NB2Slides is useful for constructing the structure of the slides, with an average rating of 6.2 ($\sigma$=1.47). 
Participants commented that the generated structure is in good practice: ``\textit{I can easily follow its structure and incorporate my ideas. It guides me to make a very complete presentation}'' (U09); ``\textit{it kind of structure the way I would present my work}'' (U11).
}

\begin{figure*}
  \centering
  \includegraphics[width=0.85\linewidth]{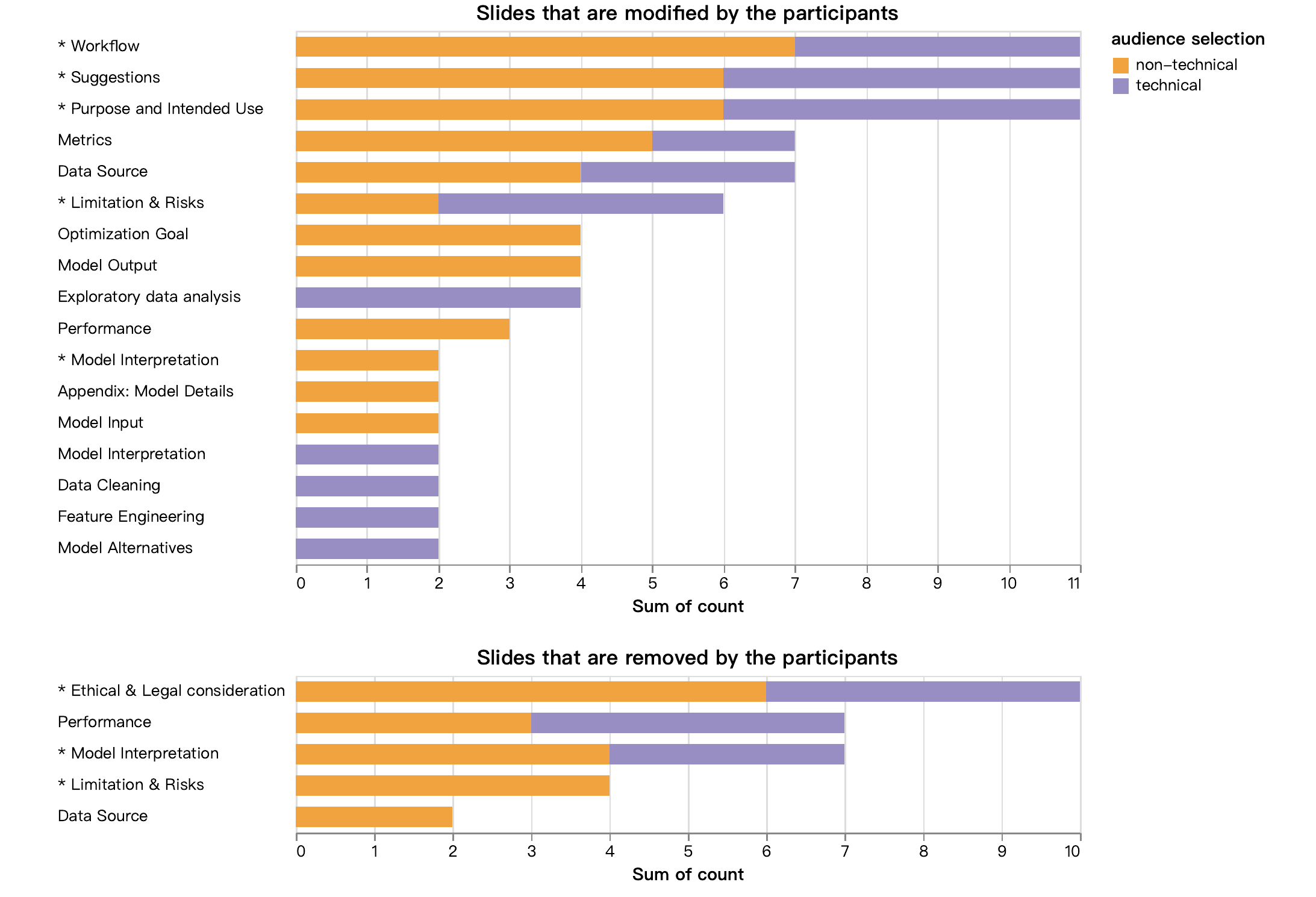}
  \caption{\label{fig:ranking} Sections that participants modified or removed, organized by the participants' configuration whether the audience background has technical or non-technical background. ``*'' denotes the content of this section is generated using example-based approach. 
  }
\end{figure*}

Second, participants thought the automated generated contents in the slides can save their time and reduce the complexity for creating slides. 
NB2Slides generated a 15-page long slide deck (excluding the first and the last slide) for participants regardless of the input parameters. The automated generated slides take up 8 pages. By inspecting the final slides participants submitted, we found that on average, the participants created slides with 12.8 ($\sigma$=2.9) pages. 
As shown in Fig.~\ref{fig:ranking}, sections in the slides that are removed by the participants are mostly some example-based prompts, e.g., \textit{Ethical \& Legal Consideration} and \textit{Model Interpretation}, indicating that most automated generated sections are either modified or directly accepted by the participants.

In the interviews, participants also reported that most automated generated contents are accurate and trustworthy: ``\textit{it's an easy summary of my notebook}'' (U01); ``\textit{At first, I checked the slides, but after two or three, I found the information is quite accurate. So I trust in the later process.}'' (U06).
U03 acknowledged NB2Slides is ``\textit{good in terms of efficiency}'' and he consider NB2Slides would be helpful to in some urgent cases: ``\textit{If you want to complete a presentation within an hour, such as 20 minutes or a shorter time, this tool is very good}''.
U08 commented \textit{``The good thing is that there's a lot of mundane activity that goes on, which is basically enough copying stuff automatically instead of the person copying it''}. 

\begin{figure*}
  \centering
  \includegraphics[width=0.8\linewidth]{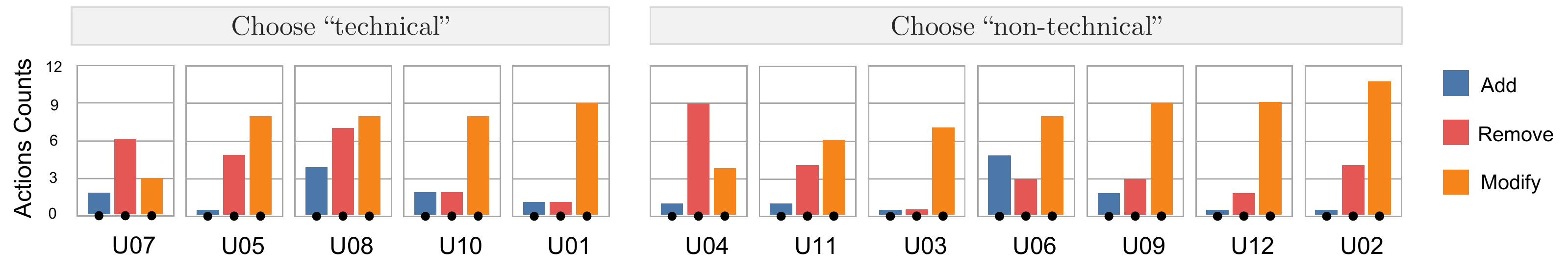}
  \caption{\label{fig:actions} \rr{
  The number of actions that participants performed regarding adding a slide, modifying an existing slide, and removing a slide. Subplots in each group are ranked by the number of modifications.
  }}
\end{figure*}

Third, participants found that manipulating the parameters like the audience background helps them to adjust slides for different audiences.
As shown in the post-study questionnaire, participants perceived positive about the usefulness of NB2Slides in customizing the slides for different audiences (mean=5.1, $\sigma$=1.6).
In particular, participants appreciate the \textit{Appendix} section that are generated with the ``non-technical'' option.
\begin{quote}
``
This actually got surprised. It generates more details into the appendix decks, so I think this is a good design. You can just talk about the results or the model to the business person and if the technical people want to know more details, then you can talk about the appendix.
''-U02
\end{quote}


\subsubsection{NB2Slides is easy to use}
From the post-study questionnaire, we found that participants felt NB2Slides easy to use, with an average rating of 6.2 ($\sigma$=0.83).
Besides the automation ability mentioned above, NB2Slides provide slides with example-based prompts. Most participants acknowledged the functionality of examples:
\begin{quote}
``Most of the words or sentences inside of these sides are placeholders, but I still found them very helpful in structuring the chain of thoughts in this presentation''-U12
\end{quote}
Participants also considered not automatically generating contents for these slides is a correct decision: 
\begin{quote}
``There's certain things that you shouldn't be automatically generating, like ethics concerns,  limitations. I like that you included those slides and ask the user to fill them out''-U07
\end{quote}

Beyond the generated slides, NB2Slides provides an interface for participants to explore and edit the slides. 
Participants gave high ratings on using NB2Slides to locate (mean=5.5, $\sigma$=1.7) and gather (mean=5.7, $\sigma$=1.6) useful information from the notebook to the slides. 
Participants in interviews gave explanations on how UI design like \textit{notebook overview} benefits the task. U03 commented: 
``\textit{If I want to see the effect of training, I don’t need to read the entire notebook. I can directly press the slide, and retrieve its corresponding information through the outline}[\textit{notebook overview}]''.
\textit{Notebook Overview} also provides explanations on the generated slides via interactive linking. Participants considered the provided explanations are informative. 
U02 commented: ``\textit{By looking at the the highlighted part I can go directly to the section}[of the notebook] \textit{to see what this slide is talking about}''.

\subsubsection{Adaptation of user behavior to automation} \label{sec:adaptation}

\input{tables/modification}

NB2Slides provides rich interactions for users to customize the generated slides.
The first is adjusting the parameters for generating slides.
We found that 10 out of 12 participants generated slides using NB2Slides in one shot.
U07 and U11 generated slides three times. Both of them first chose parameters based on their preference, then tried other generation settings, and lastly, regenerated the slides using the first-time set.
When choosing the parameters for generating slides, five participants chose ``technical'' as the audience background, and seven chose ``non-technical''. 
Fig.~\ref{fig:actions} presents these two groups of participants and the actions they performed during the study.

\rr{
In general, participants in both groups were primarily working on modifying the generated slides.
They were less likely to remove the generated slides and/or add new slides.
Such behaviors show that the generated slides serve as drafts for users so they do not need to work from scratch.
There are exceptional users: U07 and U04 removed many slides.
U07, who chose ``technical setting'' as the audience background, mainly removed the slides with example-based prompts (4 out of 6) and added slides regarding model details (e.g., ``list of features''), showing the focus of his presentation is around technical stuff.
On the contrary, targeting a ``non-technical'' audience, U04 removed slides about technical details, which were already located in the slide deck's \textit{appendix} section.
These examples show the diverse personal preferences of DS workers.
}

From the interviews, we found that some participants chose ``techincal'' because they desired to show the complexity of the DS work, despite we told them there are non-technical audiences. U07, who generated the slides three times and finally chose the ``technical'' setting, said: ``\textit{You shouldn't hide the complicated nature of what you did, because then they might think that you didn't actually do any work}''.
Besides, some participants chose ``technical'' because of their preferences for working with automation. U01 commented that ``\textit{the idea here was to get maximum information onto the slides and then start filtering if I cannot find information useful}''.
Such a do-by-filtering style does not fit everybody. Some other participants chose an incremental way that generates the ``non-technical'' slides and then adds some critical stuff. 
In that case, U09 commented that the ``\textit{Appendix}'' can be helpful: ``\textit{The appendix can serve as like a area for additional data that I can copy from and add items into the previous slide}''. 

We also retrieved some insights by analyzing the system logs.  As shown in Fig.~\ref{fig:ranking}, the participants who chose ``technical'' audience background have a higher tendency to modify the automated generated slides that concerning technical details like \textit{EDA} etc. 
For participants who chose ``non-technical'', they toke less times modifying details like \textit{EDA}, \textit{Feature Engineering} and \textit{Data Cleaning}, which were sections under the \textit{Appendix}. 
But they modified slides like \textit{Metrics}, \textit{Data Source}, \textit{Optimization Goal} and \textit{Model Output}, that are before the \textit{Appendix}. 
By looking into their final slides, we found they were adding more context-related words into the generated sentences to make them more understandable(Table.~\ref{tab:compare}). For example, NB2Slides generates a sentence, ``Fit the model and predict accuracy'', for \textit{Model Output}. U02 and U11 both added ``\textit{wine}'' as the scope to make it more clear to the audience.


\subsubsection{Future improvements}
\label{sec:futureimprove}
Participants also have concerns on NB2Slides and expect it can be improved in the future.
First, participants reported that the automated generated slides are not always accurate and expect the quality of the generated slides can be improved.
U02 commented that the inaccuracy ``\textit{will add a lot of noise to the user}''.
Besides, some participants hope NB2Slides can find more technical stories from the codes. 
P05 said, ``\textit{sometimes I may have some try outs and then I have to put more trial-and-errors thing. I don't see this system have that purpose now}''.
What's more, \nobreak participants expected NB2Slides can accept more complex input and produce more interactive artifacts. U08 mentioned the hardship of the code quality: ``\textit{in practice the workflows can be different, can be very messy}''. U11 reported the demands for interactive visualizations, ``\textit{If people are specializing in data visualization, like a kind of present to executives, they won't feel enough to put static charts in PowerPoint. They will build a dashboard, make it really interactive and pretty}''.





\subsection{Human-AI Collaboration for Slide Creation in Data Science}
The participants in our study worked closely with the AI assistant to create presentation slides. 
NB2Slides applied automated \nobreak methods to draft the slides, and the participants customized the slides on their needs and styles. 
The results in Sec.~\ref{sec:adaptation} have discussed the role of human in this collaboration process. In this section, based on the above study process, we present our results of the role of automation in creating slides and participants' attitudes to full automation.

\subsubsection{The role of automation in DS works} 
As shown in Sec.~\ref{sec:result1}, participants believed NB2Slides's automation provide them with a clear slide structure and accurately generated contents in most cases. 
However, as shown in Fig.~\ref{fig:ranking}, automated generated slides are still modified a lot by the participants.
By further looking at Table.~\ref{tab:compare}, we can found automation has different functions for different modifications to the automated generated sentences. 
The most straightforward modification is deleting unnecessary words from the generated sentences like U09 did for the sentence of \textit{Optimization Goal}.
Under such modifications, the meanings of the generated sentences are preserved.
Another modification is adding or modifying some words in the generated sentences like U01 and U02 did for the sentence under \textit{EDA}. 
Automation works like hooks to induce participants to write some relevant content.
The most radical modification is to delete the generated sentences completely and rewrote them by participants themselves as U07 did for the sentence under \textit{Model Output}.
The generated sentences are more like placeholders.

In summary, the automation generated content is not in a ready-to-present shape for participants to directly use in a presentation. The generated sentences serve as raw material and require further human improvement:

\begin{quote}
``I feel like it's more like documentation[...]The model did a quite good job in doing those summarizing because those are standard procedures like alternatives are listing all the models here. But by the time it's coming to like, you need to add some like business context to the output or summarizing''-U11
\end{quote}

\subsubsection{Full automation or human-AI collaboration in the future of DS work?}
We asked the participants to envision a future system that could fully automated generate the slides during the interviews. 
Four participants argued such full automation is impossible to achieve. 
In particular, they believed it is infeasible to generate some slides that NB2Slides provided examples for, e.g., \textit{Introduction} and \textit{Conclusion}.
Moreover, all but one participant considered they still need human intervention. P04, as the only exception, added a requirement to the full automation: ``\textit{it's comparable to my own way of making slides}''. Other participants suggested at least they need time to recheck the generated results: ``\textit{I will check my slides with more than two times. I think if it is the full generated slide, I also need a double-check or more}''(U06)

We stepped further to get deeper insights on why the participants did not trust the full automation and require human intervention. One common reason is that they thought there was not enough information, like business context, in the notebook to generate reliable slides for presentation use.

\begin{quote}
`` [...] Because the notebook itself does not provide those business context. It's like we cannot bring \textbf{the model} and \textbf{the brilliant mind} behind it, right?''-U11
\end{quote}

What's more, participants suspected whether the automated results could match the human-level, ``\textit{innovation and complete logic}''(U03). Sometimes, the ultimate goal of slide creation is not to do it fast but do it right:

\begin{quote}
``Let's say you spend a month working on a model and then at the end, you will be presenting it, I think you would probably want to take your time to spend a day or two doing it more manually because You want to \textbf{get the stories out} there''-U07

\end{quote}


%% file: tables/modification.tex
\renewcommand{\arraystretch}{1.2}
\begin{table*}
    \centering
    \begin{tabular}{@{}lp{6cm}p{6cm}@{}} \toprule
     Section & Generated by NB2Slide                                                          & Edited by Participants \\ \midrule
     \multirow{2}{*}{EDA}               & \multirow{2}{*}{\parbox{6cm}{Create a heatmap using the correlation matrix}} & \parbox{6cm}{Create a heatmap using the correlation matrix \textit{between different features and labels}~(U02)} \\  \cmidrule{3-3}
                                        &                                                                                & \parbox{6cm}{Heatmap \textit{of wine sample indices for each column}~(U01)}\\ \midrule
     Metrics                            & \parbox{6cm}{Compute the F1 score of a model}                                & \parbox{6cm}{F1 score: \textit{How well the model is handling false positive and false negative examples}~(U12)} \\ \midrule
     \multirow{2}{*}{Optimization Goal} & \multirow{2}{*}{\parbox{6cm}{Compute the cross validation score of models}}  & \parbox{6cm}{cross validation scores~(U09)} \\ \cmidrule{3-3}
                                        &                                                                                & \parbox{6cm}{\textit{Maximize the wine's prediction accuracy against the scoring of human experts}~(U12)} \\ \midrule
     \multirow{3}{*}{Model Output}      & \multirow{3}{*}{\parbox{6cm}{Fit the model and predict accuracy}}            & \parbox{6cm}{Fit the model and predict \textit{wine} quality~(U02)}  \\ \cmidrule{3-3}
                                        &                                                                              & \parbox{6cm}{predict \textit{the quality of wine (Good vs. Bad)}~(U11)} \\ \cmidrule{3-3}
                                        &                                                                              & \parbox{6cm}{\textit{Use an additional 20\% of the data for evaluating final metrics}~(U07)} \\
    \bottomrule
    \end{tabular}
    \caption{Examples of how participants edited the generated sentences. }
    \label{tab:compare}
\end{table*}

%% file: sections/07_Discussion.tex
\subsection{Human-AI Collaboration on Presentation Slides Creation in DS}
Motivated by literature and a formative study, we designed and implemented NB2Slides, an AI-assisted slides creation system for data science work.
We conducted a follow-up user study to evaluate the effectiveness of NB2Slides and the potential of automation in supporting DS workers in creating presentation slides.
The user evaluation results reveal that NB2Slides can support the slides creation process from two aspects: first, the slides generated by the deep learning model are perceived as good ``raw material''; second, the example-based prompts, the user interface and the interactive linking facilitate the editing and refinement to the slides.
Overall, the participants found that collaborating with the AI component in NB2Slides system improved their efficiency and reduced the difficulty of locating and organizing the information from the notebook, as well as customizing the presentation slides to the audience. 

We argue that the process of creating slides in our user study can be viewed as a Human-AI collaboration case, in which AI deals with complex but mechanical tasks, and humans take the draft of AI for further customization to fit their scenario.
Our study belongs to a large body of prior studies discussing the collaboration relationship between human and AI in DS~\cite{wang_collaborative_ds_2019, wang2021themisto, wang2021autods, cai2019human}.

The participants also reported that they were more in favor of having a human intervention to the final slides instead of full automation.
We found two strong reasons for DS workers to prefer the Human-AI collaboration paradigm in creating presentation slides.
The first is about the outside ``world'' of the DS work. The codes in the notebook can not capture DS workers' understanding of the implication and business value of the models, in another word, ``\textit{the brilliant minds}'' behind the notebook.  
Taking codes as input, AI can not generate sentences concerning the complex social and business world.
The second is about the inside ``secrets'' of the DS work. 
The current AI technology can only summarize what DS workers do but not why they do it in this way. As the participant mentioned, AI can not replace humans in ``\textit{getting the stories}'' our of the notebooks.

Currently, NB2Slides relies on the final completed notebook, which is the outcome of the Model Building \& Training stage in Fig.~\ref{fig:AILifecycle}, to generate slides.
We envisioned that, by having more data from the DS/ML lifecyle, NB2Slides can bring Human-AI collaboration to broader application scenarios in data science.
For example, by recording the working process of the Feature Engineering and Model Building \& Training stages(Fig.~\ref{fig:AILifecycle}), NB2Slides can gain more information from the intrinsic and iterative activities DS workers performed, as well as extract and summarize the complex process into a raw material, like what it does now. 
Based on the raw material, DS workers can further elaborate and tell their stories out of the codes.

\rr{
\subsection{Integrating AI-Assisted Presentation Creation in DS Worker's Workflow}
Presentation creation is critical for DS workers to turn their complex technical work into valuable outputs that stakeholders can \nobreak understand and trust. We observed in our formative study that when creating a presentation based on a computational notebook, DS workers had to switch between the notebook and the slide authoring tool frequently. They also faced challenges locating information, organizing a story, and customizing the slides for a target audience. NB2Slides is designed to address these challenges and fit into DS workers’ practice.

We built NB2Slides inside Jupyter Lab, a familiar exploring and modeling environment for DS workers.
NB2Slides reduces the context switching costs between the notebook and the slides for DS workers by
1) laying out notebooks and slides side-by-side; 2) coordinating interaction between these two artifacts; and 3) automatically locating information in the notebook and summarizing them to the slides. 
Besides, a notable behavior of DS workers in creating presentations is that they tend to reuse their past presentation slides and adapt the content to the current work ~\cite{piorkowski2021ai}.
The function of NB2Slides can also be viewed from the perspective of ``reuse'': NB2Slides is reusing the good practice of DS presentation (our validated template) by automatically filling in the relevant content from the input notebook.
The automation helps with adapting slides contents and lowers DS workers' loads about locating, organizing, and customizing information. 

Considering the AI technology used behind NB2Slides, several potential risks exist for DS workers.
First, the uncertainty of the slides generated from messy input notebooks may cost extra time for DS workers to clean the notebook and refine the generated slides.
Cleaning the notebook is not a standalone task for creating a presentation using NB2Slides. 
Instead, DS workers often need to refactor their notebooks for sharing or collaborating purposes~\cite{rule2018exploration, head2019managing}, which can be supported by several existing tools that are also built in the Jupyter Lab~\cite{drosos2020wrex, head2019managing, weinman2021fork}.

Second, with long-term usage of NB2Slides, DS workers may overtrust and overly rely on the automated solution, making DS workers less prepared for storytelling and answering questions. We believe such risk is minimal as NB2Slides is designed to focus on human-AI collaboration instead of providing a full automation solution. We observed in the evaluation study that DS workers actively refine the generated slides, which suggests that the generated slides can serve only as a rough draft to save human effort but not to eliminate human intelligence. But the future study is needed to explore the potential overreliance of AI topic further.


}

\subsection{Design Implications}

\subsubsection{Context-Driven Presentation Creation.}
One pain point we found in the formative study and the user study of NB2Slides is how to prepare presentations for different audiences and under different scenarios. NB2Slides allow users to control the generation of slides by specifying the audience's technical level.
Although participants recognized that it is a good start, they still considered it to be insufficient to characterize all the nuances about potential audience. 
Participants also suggested that giving a five-minute one-to-one presentation versus a twenty-minute public presentation would need quite different sets of slides.
Hence, we recognize that more works need to be done in terms of distinguishing different audiences and different scenario to help data scientists better tailor their presentations.
We envision that future AI-assisted slides creation could provide more fine-grained context-related control for DS workers to drive the auto-generation of slides.
To achieve this goal, researchers may conduct empirical studies to understand the needs of various audience involved in a data science workflow and possible forms of presentation under different communication scenarios. 

\subsubsection{Towards Better Human-AI Collaboration}
Currently, users can create a slide deck using NB2Slide in two phases: generation and customization. AI mainly participates in the ``generation'' phase. However, as stated by U09, ``\textit{It will be difficult to use the system if everything is done in one shot}''. 
Users expect to have more interaction with AI. To this end, a number of existing approaches can be integrated to further improve the NB2Slide system.
For example, when the user adds a new slide and enters the title, the model can immediately inference what cells in the notebook are related to the title. 
NB2Slides can then link these cells to this new slide and even recommend bullet points summarizing the cell contents to the user. 
Also, after the user modifies several slides, the model can learn from the user's historical behavior recorded in the log data and provide suggestions on how to refine future slides in a similar way (e.g., they prefer not to see particular slides, or prefer to use certain languages).

\subsubsection{Multi-Granularity Explanations for Automation}
By dynamic linking the generated slides with the cells in the notebook, NB2Slides explains how it automatically generate some slides from the notebook. These post-hoc explanations~\cite{lipton2018mythos} are welcomed by participants. However, they also suggested that more granularities of explanations are desirable. On the one hand, while current explanation stays at the slide-level, U04 expects explanations about how the configuration parameters (e.g., the audience background) would affect the slides at the section-level, ``\textit{I don't know whether the whole structure of the slides will be different if I am playing with these parameters}''.  On the other hand, U02 hopes to have some explanations at the sentence-level, ``\textit{It will be interesting if you can tell more about, for instance, how this sentence is generated}''. Based on this feedback, we concluded that multi-granularity explanations are needed for automation on artifacts with a hierarchical structure, e.g., slides.


\subsection{Limitation and Future Work}
NB2Slides as a proof-of-concept system has several limitations. It currently requires high-quality notebooks as inputs. But as noted by U08 in the evaluation study, ``\textit{in practice, the workflows can be different, can be very messy}''. NB2Slides may fail to generate reasonable slides with unorganized notebooks. Hence, we hope to improve the performance of NB2Slide by integrating existing works on augmenting the computational notebook~\cite{head2019managing, kery2019towards}. 
Also, the results of related cell extraction in NB2Slides are far from perfect. One possible reason is that the sentence embedding model we used (i.e., SimCSE~\cite{gao2021simcse}) is not pre-trained on a DS-related corpus. The results can be further improved by fine tuning the model.

Our user study also have limitations. First, we did not conduct a comparative experiment study. It is valuable to understand the usefulness of NB2Slide by comparing it with some baselines, for example, creating slides with the notebook and the Powerpoint application side-by-side (as we did in the formative study). Second, the participants used NB2Slides for one notebook provided by us. Due to the unfamiliarity with the code, participants may behave differently in comparison to how they would create a slides for their own code.  
Regardless, we argue that our participants are experienced and the notebook is fairly simple to understand, it should not be a huge burden for our participants to understand the notebook. 
Thus, the findings and insights from this study is still insightful.
Future work is needed to observe a long-term deployment to see if the benefits from this AI-assisted slides creation system persists after a while.


%% file: sections/08_Conclusion.tex
In this paper, we present NB2Slides, an AI-assisted presentation slides creation system that facilitates DS workers to locate information, organize stories, and customize slides for presenting data science work. The design of NB2Slides system is motivated by literature and a formative study with seven DS workers. NB2Slide can take customized user configurations to guide the deep-learning models to generate draft presentation slides . Users can also explore and refine the slides using NB2Slides. The follow-up user evaluation confirmed that NB2Slides improves DS workers' efficiency in creating presentation slides, and promotes the collaboration between human and AI.

%% file: sections/09_Appendix.tex
\newpage
\section{Four Examples of Generated Slides from Kaggle Notebooks and Error Analysis}

We tested NB2Slides on two kinds of notebooks:
\begin{itemize}
    \item Study material from previous works~\cite{wang2021themisto, head2019managing} related to computational notebooks. 
    \item Randomly sampled Kaggle notebooks. 
\end{itemize}

All the testings can be found in the supplementary material. Here, we show some representative examples.
Overall, NB2Slides can successfully generate presentation slides for most notebooks (e.g., the two notebooks in Table~\ref{tab:example1}) satisfying the three input requirements (Sec~\ref{sec:inputs}).
There are also some less successful cases (e.g., the two examples in Table~\ref{tab:example2}), most errors in which can be categorized into the following two kinds:

\begin{enumerate}
    \item Incorrect information locating. NB2Slides leverages language model to locate relevant cells in an input notebook for a slide section. However, some cells can be similar in comments and codes but differ greatly regarding real purposes. For example, the codes of visualization for EDA and the codes of visualization for model evaluation can be very similar. Thus NB2Slides took the later one for generating slides about EDA for N04 in Table~\ref{tab:example2}.
    \item Ineffective code summarization. NB2Slides uses a pre-trained model (codeTrans~\cite{elnaggar2021codetrans}) to summarize python codes. But some generated sentences are not ready for presentation. For example, ``Score with cv\_rmse'' is generated for the ``Model Performance'' section of S01 in Table~\ref{tab:example1}. A more desirable sentence can be ``RMSE scores with cross-validation''. 
\end{enumerate}

We believe more detailed documentation can help to prevent both kinds of errors. 
Fine-tuning the code summarization model over DS-related codes can also benefit to prevent the second kind of error.

Worth mentioning, the interaction enabled by the interface of NB2Slides can help users handle these errors.
Using interactive linking, they can know which cell does NB2Slides use for generating a specific point.
They can judge if the AI model locates the correct information for generation, thus determining whether to preserve the point (first kind of error).
If they find the sentence of the point hard to read, they can also use the interactive linking to quickly find the source code to refine the sentence (second kind of error).

\newpage

\begin{table*}[t]
\centering%
\begin{tabular}{@{}p{4pc}|p{15pc}|p{13pc}@{}}%
\toprule%
\textbf{Slides ID} & S01 & S02\\ \midrule%
{}{\textbf{Input}} &{} &{}\\%
{\textbf{Notebook}} & House Price Prediction & COVID-19 Prediction{}\\ \midrule%
\textbf{Notebook} &&{}\\%
\textbf{Source} & Kaggle~\footnote{\url{https://www.kaggle.com/alfredmaboa/advanced-regression-techniques-regularization}} & Study Material~\cite{wang2021themisto}  \\
        \textbf{IR1}    & {\includegraphics[scale=0.05]{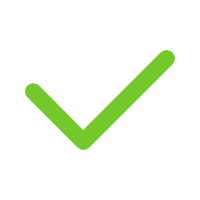}} &  {\includegraphics[scale=0.05]{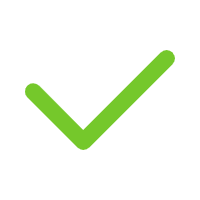}} \\
        \textbf{IR2}    & {\includegraphics[scale=0.05]{figures/icons/correct.png}}  &  {\includegraphics[scale=0.05]{figures/icons/correct.png}} \\
        \textbf{IR3}    & {\includegraphics[scale=0.05]{figures/icons/correct.png}} &   {\includegraphics[scale=0.05]{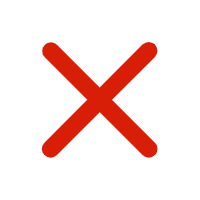}}~ (lack Feature Engineering) \cr  \midrule
&  \textbf{EDA}~{\includegraphics[scale=0.05]{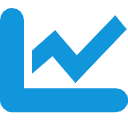}}&\textbf{EDA}\cr
\smash{\raisebox{-2pc}{\textbf{Generated}}} &{\includegraphics[scale=0.05]{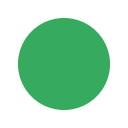}}~ Plot the distribution of a new sale price& {\includegraphics[scale=0.05]{figures/icons/good.png}}~ Read the CSV files and return \cr
\smash{\raisebox{-2pc}{\textbf{Slides}}}&{\includegraphics[scale=0.05]{figures/icons/good.png}}~ Plots the distribution of numeric features &a DataFrame of 5 rows.\cr
&as a boxplot &{\includegraphics[scale=0.05]{figures/icons/good.png}}~ Generate descriptive statistics\cr
\smash{\raisebox{-1pc}{\textbf{Contents}}}&{\includegraphics[scale=0.05]{figures/icons/good.png}}~ Plots sales prices for numeric features \cr \cline{2-3}
&   \textbf{Data Cleaning} & \textbf{Data Cleaning}\cr
&{\includegraphics[scale=0.05]{figures/icons/good.png}}~ Returns a list of the most likely \hfill\break missing values in each column. & {\includegraphics[scale=0.05]{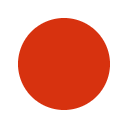}}~ Read the CSV files and return \hfill\break a DataFrame of 5 rows.\cr
&{\includegraphics[scale=0.05]{figures/icons/good.png}}~ This function replaces all missing data \hfill\break with NaN so that the number of \hfill\break non-numeric predictors is not used &{\includegraphics[scale=0.05]{figures/icons/good.png}}~ Check for missing values \hfill\break in the train data\cr
&{\includegraphics[scale=0.05]{figures/icons/good.png}}~ Returns a dataframe of the percentage of\hfill\break missing values in each feature  \cr \cline{2-3}
&  \textbf{Feature Engineering} & \textbf{Feature Engineering} \cr
&{\includegraphics[scale=0.05]{figures/icons/good.png}}~ Adding new features to the feature list &NULL \cr 
&{\includegraphics[scale=0.05]{figures/icons/good.png}}~ Encode the finalized features into a dataframe\cr \cline{2-3}
& \textbf{Model Performance}~{\includegraphics[scale=0.05]{figures/icons/plot.png}} &\textbf{Model Performance}\cr
&{\includegraphics[scale=0.05]{figures/icons/good.png}}~ Root Mean Square Error scoring function &{\includegraphics[scale=0.05]{figures/icons/good.png}}~ Predicts the case for each test in x\_test\cr
&{\includegraphics[scale=0.05]{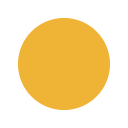}}~ Score with cv\_rmse&\cr
&{\includegraphics[scale=0.05]{figures/icons/good.png}}~ Plots the model performance&\cr \cline{2-3}
& \textbf{Model Details} &\textbf{Model Details}\cr
&{\includegraphics[scale=0.05]{figures/icons/good.png}}~ Root Mean Square Error scoring function &{\includegraphics[scale=0.05]{figures/icons/good.png}}~ Create the target and test data arrays\cr
            &{\includegraphics[scale=0.05]{figures/icons/medium.png}}~ Score with cv\_rmse &{\includegraphics[scale=0.05]{figures/icons/good.png}}~ Random Forest Classifier\cr
&{\includegraphics[scale=0.05]{figures/icons/good.png}}~ Compute RMSLE of fitted models &{\includegraphics[scale=0.05]{figures/icons/good.png}}~ Predicts the case for each test in x\_test.\cr
&{\includegraphics[scale=0.05]{figures/icons/good.png}}~ Plots the [ \textit{best} ] model performance &\cr
&{\includegraphics[scale=0.05]{figures/icons/medium.png}}~ Predict the modeling for a sample  &\cr
    \bottomrule
    \end{tabular}
    \caption{Two successful examples with input notebook's meta information and the generated slides  contents. For each bullet point, a dot is positioned before it to represent its quality: {\protect\includegraphics[scale=0.05]{figures/icons/bad.png}}~ stands for incorrect information locating; {\protect\includegraphics[scale=0.05]{figures/icons/medium.png}}~ shows ineffective code summarization; {\protect\includegraphics[scale=0.05]{figures/icons/good.png}}~ shows the point is good for usage. Additionally, {\protect\includegraphics[scale=0.05]{figures/icons/plot.png}} indicates output plots are carried with that section in the generated slides; ``NULL'' means the system does not generate any point.}
    \label{tab:example1}
\end{table*}

\begin{table*}[t]
    \centering
    \begin{tabular}{@{}p{4pc}|p{16pc}|p{16pc}@{}} 
    \toprule
        \textbf{Slides ID} & S03 & S04 \cr \midrule
        \textbf{Input} \hfill\break \textbf{Notebook} & Happiness Starting Point & Ventilator Pressure Prediction \cr \midrule
        \textbf{Notebook}\hfill\break  \textbf{Source} & Study Material~\cite{head2019managing} & Kaggle~\footnote{\url{https://www.kaggle.com/mst8823/19th-place-best-single-model-resbilstm}} \cr
        \textbf{IR1}  &  {\includegraphics[scale=0.05]{figures/icons/correct.png}} & {\includegraphics[scale=0.05]{figures/icons/correct.png}} \cr
        \textbf{IR2}  &  {\includegraphics[scale=0.05]{figures/icons/wrong.png}} & {\includegraphics[scale=0.05]{figures/icons/wrong.png}} \cr
        \textbf{IR3}  &  {\includegraphics[scale=0.05]{figures/icons/wrong.png}}~ (lack Feature Engineering)  & {\includegraphics[scale=0.05]{figures/icons/wrong.png}}~ (lack Data Cleaning) \cr  \midrule
&\textbf{EDA}~{\includegraphics[scale=0.05]{figures/icons/plot.png}} & \textbf{EDA}~{\includegraphics[scale=0.05]{figures/icons/plot.png}} \cr
\smash{\raisebox{-1pc}{\textbf{Generated}}}           &{\includegraphics[scale=0.05]{figures/icons/good.png}}~ Figures of corruption and happiness. &            {\includegraphics[scale=0.05]{figures/icons/bad.png}}~ Runs the CV model on both sets of features and\hfill\break  returns a dataframe with all relevant metrics\cr
\textbf{Slides}           &{\includegraphics[scale=0.05]{figures/icons/good.png}}~ Plots the Happiness scores for 2015 and 2017. &{\includegraphics[scale=0.05]{figures/icons/bad.png}}~ Plot the oof score for each breath\cr
\textbf{Contents}             &{\includegraphics[scale=0.05]{figures/icons/good.png}}~ Plots the Happiness scores for each year &{\includegraphics[scale=0.05]{figures/icons/bad.png}}~ Displays the oof score of a train set \cr \cline{2-3}
       & \textbf{Data Cleaning} & \textbf{Data Cleaning}\cr 
        &    NULL&{\includegraphics[scale=0.05]{figures/icons/bad.png}}~ This environment is Kaggle Kernel.\cr
          &&{\includegraphics[scale=0.05]{figures/icons/bad.png}}~ This function takes a csv file and create\hfill\break  pandas dataframes with the training test \& sample\hfill\break  submission files. \cr   \cline{2-3}
            & \textbf{Feature Engineering} &\textbf{Feature Engineering}\cr
            &NULL &{\includegraphics[scale=0.05]{figures/icons/bad.png}}~ Experiment configuration\cr 
            &&{\includegraphics[scale=0.05]{figures/icons/good.png}}~ Masks the features of all\hfill\break  animals and groups them by time step
        \cr \cline{2-3}
        & \textbf{Model Performance}~{\includegraphics[scale=0.05]{figures/icons/plot.png}} &\textbf{Model Performance}~{\includegraphics[scale=0.05]{figures/icons/plot.png}}\cr
&            {\includegraphics[scale=0.05]{figures/icons/bad.png}}~ Splits the happiness average score by region\hfill\break  into groups of regions and scores. &{\includegraphics[scale=0.05]{figures/icons/good.png}}~ Runs the CV model on both sets of features and\hfill\break  returns a dataframe with all relevant metrics\cr
&          {\includegraphics[scale=0.05]{figures/icons/bad.png}}~ Skin - based plot of happiness rank. &            {\includegraphics[scale=0.05]{figures/icons/good.png}}~ Plot the oof score for each breath\cr
&{\includegraphics[scale=0.05]{figures/icons/good.png}}~ Approximates the Happiness scores of each species& {\includegraphics[scale=0.05]{figures/icons/good.png}}~ Displays the oof score of a train set \cr \cline{2-3}
        & 
            \textbf{Model Details} &\textbf{Model Details} \cr
            &{\includegraphics[scale=0.05]{figures/icons/bad.png}}~ Attach a text label above each bar\hfill\break  displaying its height. &{\includegraphics[scale=0.05]{figures/icons/good.png}}~ Build a Keras model with the masked\hfill\break mean absolute error.\cr
            &{\includegraphics[scale=0.05]{figures/icons/good.png}}~ Simple Regression using simple linear regression.&{\includegraphics[scale=0.05]{figures/icons/good.png}}~ Loads the dataset for training and testing\cr
            &&{\includegraphics[scale=0.05]{figures/icons/medium.png}}~ Plots the OOF score for\hfill\break  each R - C in a train set\cr
            &&{\includegraphics[scale=0.05]{figures/icons/good.png}}~ Displays the oof score of a train set\cr
            &&{\includegraphics[scale=0.05]{figures/icons/good.png}}~ Plot bad predictions by R \& C \cr
    \bottomrule
    \end{tabular}
    \caption{
        \protect\rr{
        Two failure examples with input notebook's meta information and the generated slides  contents. Symbols are with the same meanings as Table.\ref{tab:example1}.
        }
    }
    \label{tab:example2}
\end{table*}